\documentclass[myepj-spec]{mySvjour}
\usepackage{graphicx}
\usepackage{hyperref}
\usepackage{color}
\usepackage{xcolor}
\usepackage{xspace}
\usepackage[square,sort&compress,numbers]{natbib}   
\usepackage{verbatim}
\usepackage{amsmath,amssymb,amsfonts} 
\usepackage{tabularx}
\usepackage{multicol}
\usepackage{footnote}
\usepackage{listings}
\usepackage{orcidlink}
\usepackage{diagbox}
\usepackage[switch, modulo]{lineno}

\usepackage{lmodern,bm}   
\usepackage{soul}
\usepackage{booktabs}

\usepackage[T1]{sansmath} 
\SetMathAlphabet{\mathsfbf}{sans}{\sansmathencoding}{\sfdefault}{bx}{sl}
\usepackage{etoolbox}
\AtBeginEnvironment{sansmath}{}{}{}

\usepackage{lipsum}

\newcommand{\GravNet}{\textsc{GravNet\,}}

\definecolor{darkblue1}{rgb}{0,0,.2}
\definecolor{darkblue}{rgb}{0,0,.2}
\definecolor{darkred}{rgb}{0.5,0,0}
\definecolor{darkgreen}{rgb}{0.25, 0.47, 0}

\pagecolor{white} 
\color{black}     
\hypersetup{breaklinks=true, 
	colorlinks=true, 
	linkcolor=darkblue1, 
	menucolor=darkblue1, 
	urlcolor=darkblue1,
	citecolor=darkblue1,
	pdftitle={},
	pdfauthor={},
	pdfsubject={},
	pdfkeywords={},
	pdfproducer={}
}
\newcommand{\DB}[1]{{\color{blue}{DB: {#1}}}} 
\newcommand{\YG}[1]{{\color{purple}{YG: {#1}}}} 
\newcommand{\DBT}[1]{{\color{teal}{DBT: {#1}}}} 
\newcommand{\CG}[1]{{\color{red}{CG: {#1}}}} 

\newcommand{\GM}[1]{{\color{olive}{GM: {#1}}}} 

\usepackage{siunitx}

\bibstyle{plain}
\newcommand{\bi}{\begin{itemize}}
\newcommand{\ei}{\end{itemize}}

\newcommand{\ben}{\begin{enumerate}}
\newcommand{\een}{\end{enumerate}} 

\newcommand{\bt}[1]{\begin{table}[tb]\begin{tabular}{#1} \hline\hline  \\[-1.0em]}
\newcommand{\et}[2]{\hline\hline \end{tabular} \caption{#1} \label{#2} \end{table}}

\newcommand{\be}{\begin{equation}}
\newcommand{\ee}{\end{equation}}

\newcommand{\bea}{\begin{eqnarray}}
\newcommand{\eea}{\end{eqnarray}}

\newcommand{\bc}{}

\newcommand{\mev}{\ensuremath{\mathrm{\,Me\kern -0.1em V}}\xspace}
\newcommand{\gev}{\ensuremath{\mathrm{\,Ge\kern -0.1em V}}\xspace}

\begin{document}
	
	{%
		\begin{@twocolumnfalse}
			
			\begin{flushright}
				\normalsize
			\end{flushright}
			
			\vspace{-2cm}
			\title{\Large\boldmath Global detector network to search for high-frequency gravitational waves (GravNet): conceptual design}
			%

\author{
Dorian Amaral$^{1}$\orcidlink{0000-0002-1414-932X},
Diego Blas$^{1,12}$\orcidlink{0000-0003-2646-0112}, 
Yuliia Borysenkova$^{1,14}$\orcidlink{0000-0003-1040-2815},
Dmitry Budker$^{2,3,4,5}$\orcidlink{0000-0002-7356-4814}, 
Alessandro D'Elia$^{6}$,
Giorgio Dho$^{6}$,
Alejandro Díaz-Morcillo$^{11}$,
Daniele Di Gioacchino$^{6}$,
Sebastian Ellis$^{17}$,
Claudio Gatti$^{6}$, 
Benito Gimeno$^{9}$,
Jordan Gu\'e$^{1}$\orcidlink{0009-0000-4383-285X},
Stefan Horodenski$^{8}$\orcidlink{0009-0009-6848-1094},
Saarik Kalia$^{1}$,
Younggeun Kim$^{2,3}$\orcidlink{0000-0001-7297-8110},
Tom Krokotsch$^{7}$,
Tomas Kvietkauskas$^{1,14}$\orcidlink{0009-0005-3844-4488},
Adrián Lambíes-Asensio$^{9}$,
Carlo Ligi$^{6}$,
Giovanni Maccarrone$^{6}$,
Giovanni Mazzitelli$^{6}$,
Juan Monzó-Cabrera$^{11}$,
José R. Navarro-Madrid$^{11}$,
José Reina-Valero$^{10}$,
Alessio Rettaroli$^{6}$,
Kristof Schmieden$^{8}$\orcidlink{0000-0003-1978-4928}, 
Tim Schneemann$^{13}$\orcidlink{0000-0002-1460-5292}, 
Matthias Schott$^{8}$,
Simone Tocci$^{6}$,
Sandro Tomassini$^{6}$,
Oleg Tretiak$^{2,3}$\orcidlink{0000-0002-7667-2933},
Luca Visinelli$^{15,16}$\orcidlink{0000-0001-7958-8940},
Changhao Xu$^{2,3}$\\
}

\institute{\it
\inst{1} Institut de F\'{i}sica d’Altes Energies (IFAE), The Barcelona Institute of Science and Technology, Campus UAB, 08193 Bellaterra (Barcelona), Spain\\
\inst{2} PRISMA++ Cluster of Excellence, Institute of Physics, Johannes Gutenberg University, Mainz, Germany \\ 
\inst{3} Helmholtz Institute Mainz, 55099 Mainz, Germany \\
\inst{4} GSI Helmholtzzentrum für Schwerionenforschung GmbH, 64291 Darmstadt, Germany \\
\inst{5} Department of Physics, University of California, Berkeley, California 94720, USA \\
\inst{6} National Institute for Nuclear Physics, Italy\\
\inst{7} Universit\"at Hamburg,  Luruper Chaussee 149, 22761 Hamburg, Germany\\
\inst{8} Rheinische Friedrich-Wilhelms-University, Bonn, Germany \\
\inst{9} Instituto de Física Corpuscular (IFIC), CSIC-University of Valencia, Calle Catedrático Jose Beltrán Martínez, 2, 46980 Paterna (Valencia), Spain \\
\inst{10} Laboratorio Subterráneo de Canfranc, 22880 Canfranc-Estación (Huesca), Spain \\
\inst{11} Departamento de Tecnologías de la Información y las Comunicaciones, Universidad Politécnica de Cartagena, Plaza del Hospital 1, 30202 Cartagena (Murcia), Spain. \\
\inst{12} Instituci\'{o} Catalana de Recerca i Estudis Avan\c{c}ats (ICREA), Passeig Llu\'{i}s Companys 23, 08010 Barcelona, Spain\\
\inst{13} Institute of Physics, Johannes Gutenberg University, Mainz, Germany \\
\inst{14} Grup de Física Teòrica, Departament de Física,
Universitat Autònoma de Barcelona, 08193 Bellaterra (Barcelona), Spain \\
\inst{15}Dipartimento di Fisica ``E.R.\ Caianiello'', Universit\`a degli Studi di Salerno,\\
\phantom{\inst{15}}Via Giovanni Paolo II, 132 - 84084 Fisciano (SA), Italy\\
\inst{16}Istituto Nazionale di Fisica Nucleare - Gruppo Collegato di Salerno - Sezione di Napoli,\\ \phantom{\inst{16}}Via Giovanni Paolo II, 132 - 84084 Fisciano (SA), Italy\\
\inst{17}Theoretical Particle Physics and Cosmology (TPPC) Group, Department of Physics, \\King's College London, Strand, London, WC2R 2LS, UK,\\
}

            \abstract{We propose \GravNet (Global detector network to search for high-frequency gravitational waves), a novel experimental scheme enabling the search for gravitational waves in the MHz to GHz frequency range. 
            Such high-frequency gravitational waves could arise from a variety of phenomena connected to some of the most pressing and fundamental questions in modern cosmology. 
            The \GravNet concept is based on synchronous measurements of signals from multiple experimental measurement devices operating at geographically separated locations. 
            While gravitational-wave–induced signatures may be present in the signal of a single detector, distinguishing them from instrumental or environmental noise is highly challenging. 
            By analyzing correlations between signals from several distant detectors, the detection significance is substantially enhanced, while simultaneously enabling studies of the nature and origin of the gravitational-wave signal. 
            In this work, we discuss the \GravNet concept specifically in the context of cavities operated in strong magnetic fields, as these currently represent the most technically mature and experimentally advanced realization of the scheme. 
            As part of this proposal, a first demonstration experiment using a non-superconducting cavity has been performed, providing the basis for the data-analysis strategies discussed in this work. 
            Finally, we outline the prospects and future development of \GravNet as a global network for high-frequency gravitational-wave searches.}
	\maketitle
	\end{@twocolumnfalse}
}

\newpage
	
\tableofcontents

\section{Introduction}

The detection of gravitational waves (GWs) stands out as one of the most groundbreaking achievements in $21^{\rm st}$-century physics. These subtle signals have revolutionized our understanding of Nature despite having been observed only in the band from 10\,Hz to 10\,kHz scanned by ground-based interferometers~\cite{LIGOScientific:2016aoc}, and suggested as the origin of a nHz signal present in pulsar-timing arrays~\cite{NANOGrav:2023gor}. In both cases, the original proposals searching for these signals were once considered futuristic, but today's results are a testament to the foresight of their pioneers.

This remarkable accomplishment ushers in new and immediate challenges. One of the most pressing issues is expanding the sensitivity to gravitational waves across other spectral regions. The discovery potential here is extraordinary: from profound inquiries into the primordial universe, like stochastic GWs from inflation or phase transitions, to unique insights into new astrophysical phenomena, such as black holes of a variety of masses, of primordial or astrophysical origin~\cite{LISA,Aggarwal:2025noe}. Several proposals have emerged to address this challenge for frequencies below 10\,kHz. For instance, GWs from the lowest detectable frequencies of $10^{-18}$ Hz (with wavelengths on the order of the size of the visible Universe today) may impact the properties of the cosmic microwave background~\cite{Clarke:2020bil}, while space-based laser interferometers will probe the mHz band~\cite{LISA}. The next generation of ground-based observatories~\cite{Maggiore:2019uih} and pulsar-timing efforts~\cite{Janssen:2014dka}, together with other techniques ranging from large atom interferometers to astrometry, provide broad coverage of GWs from $10^{-18}$\,Hz to 10\,kHz~\cite{Renzini:2022alw}.

The Global Network of Detectors to Search for Gravitational Waves (\GravNet) aims to establish a cohesive initiative targeting similar coverage of high-frequency gravitational waves (HFGWs) in the MHz to GHz range. 
A key difference in the search for HFGWs, as compared to searches at lower frequencies, is the difficulty in producing substantial GWs above 10\,kHz in standard astrophysical scenarios~\cite{Aggarwal:2025noe,Aggarwal:2020olq,Maggiore:2007ulw}.  The reason is that gravitation, being an extremely weak interaction, requires large coherent sources (such as accelerated stars or black holes) for the efficient production of GWs. However, the forces required to induce motions that generate GWs well above 10\,kHz would destroy standard astrophysical objects \footnote{The smallest objects made of nuclear matter have a diameter in the order of 10\,km. If rotating with 1\,kHz, the speed at the equator approaches the speed of light, which shows that such objects cannot emit HFGWs. It is worth mentioning that the collisions of neutron stars may produce GWs at MHz frequencies, whose discovery would clarify new aspects of matter at high densities~\cite{Blas:2022xco}. \label{footnoteBH}}. 
By contrast, HFGWs may be generated copiously by processes related to some of the most pressing open questions about our Universe: violent phenomena in the early Universe (such as primordial phase transitions, inflation, or topological defects), thermal fluctuations of the primordial plasma and dark matter~\cite{Aggarwal:2025noe,Aggarwal:2020olq}. This shows that the absence of abundant astrophysical sources enables the search of signals from fundamental physics, motivating a vigorous exploration of the HFGW band. Furthermore, this uncharted territory may harbor surprises yet to be unveiled.  We set $\hbar = c = \epsilon_0 = 1$ unless otherwise stated.

\section{Theoretical Background} 

\subsection{Sources of  transient or coherent (and persistent) HFGWs} \label{sec:sources}

A well-motivated class of sources of high-frequency gravitational waves (HFGWs) are primordial black holes (PBHs), which may be produced in the early Universe and could constitute a significant fraction, or even all, of the dark matter~\cite{Carr:2021bzv,Green:2020jor, Carr:2020gox}. 
In addition to PBHs formed directly from large primordial density fluctuations, compact objects may also arise from the gravitational collapse of particle dark matter overdensities, such as those associated with weakly interacting massive particles near the Griest--Kamionkowski bound~\cite{LVK:2022ydq,Shandera:2018xkn}. 
For WIMP masses of order $\mathcal{O}(100~\mathrm{TeV})$, the resulting compact objects can have masses as low as $\sim 10^{-10} M_\odot$, placing them naturally in the mass range relevant for HFGW emission. 
Regardless of their origin, only sufficiently compact dark sector objects can act as efficient HFGW sources as the GW frequency emitted scales with the compactness $\mathcal{C}$ as $f_g = \mathcal{C}^{3/2}/(3\sqrt{3}\pi GM_{\rm tot})$~\cite{Giudice:2016zpa}, leaving diffuse dark MACHOs irrelevant.

A fraction of PBHs is expected to reside in binary systems that emit gravitational waves. For a quasi-circular binary of two equal-mass PBHs with individual masses $M_{\rm bh}$, combined merger mass $M_{\rm PBH}$, and orbital radius $r_b$, the gravitational-wave frequency is bounded from above by the value at the innermost stable circular orbit\footnote{At smaller orbital separations, the system enters a relativistic inspiral and merger phase, during which HFGWs are still emitted, although the simple Newtonian description breaks down.}. (ISCO),
\begin{equation}
    \omega_g \equiv 2\pi f_g \approx 14 \, \mathrm{GHz}
    \left(\frac{10^{-6} M_\odot}{M_{\rm PBH}}\right)
    \left(\frac{r_{\rm ISCO}}{r_b}\right)^{3/2},
    \label{eq:freqISCO}
\end{equation}
where $M_\odot$ is the solar mass and
$r_{\rm ISCO} \simeq 36\,\mathrm{km}\,(M_{\rm bh}/M_\odot)$ denotes the ISCO radius~\cite{Maggiore:2007ulw}. Here, $r_{\rm ISCO}/r_b \leq 1$, and Eq.~\eqref{eq:freqISCO} therefore defines the ISCO frequency $f_{\textrm{ISCO}}$ corresponding to $r_{\rm ISCO}/r_b = 1$. We use $f_{\textrm{ISCO}}$ as a reference value throughout this work. 

As a result, HFGWs can be emitted by binaries of sufficiently light PBHs one achieves GHz or larger frequencies.
The emission of gravitational waves causes the binary orbit to shrink~\cite{Maggiore:2007ulw}, leading to a characteristic chirp, i.e.\ an increase in the GW frequency with time. This provides a natural scan over resonant detector frequencies. 
While the GW frequency lies within the resonance band of an oscillator, it's response is resonantly enhanced, significantly increasing the signal. However, requiring the source to remain in the resonance band for more than $Q_l \sim 10^5$ cycles implies $M_{\rm bh} \lesssim 10^{-12} M_\odot$ at \mbox{$\omega_g \simeq 5~\mathrm{GHz}$}~\cite{Berlin:2021txa},
\begin{equation}
N_{\rm cyc} \approx 10^{-3}
\left(\frac{10^{-6} M_\odot}{M_{\rm bh}}\right)^{5/3}
\left(\frac{10^5}{Q_l}\right)
\left(\frac{1\,\mathrm{GHz}}{\omega_g}\right)^{5/3}.
\label{eq:ncycles}
\end{equation}
%
 This estimate is conservative: the effective bandwidth is larger for cavities that are not fully rung up, and heavier PBHs generate significantly stronger signals even if they do not remain resonant for the full ring-up time. We therefore defer a detailed treatment of higher PBH masses and the merger regime to future work. 

Interestingly, this mass range overlaps with a notoriously difficult-to-constrain region of PBH parameter space, in which PBHs can still constitute a significant fraction of the dark matter~\cite{Carr:2021bzv}. The expected event rate and signal strength were studied in Ref.~\cite{Franciolini:2022htd}. Assuming PBHs account for all of dark matter with a narrow mass function peaked at $M_{\rm bh} \sim 10^{-12} M_\odot$, one expects order one event per year producing a sufficiently long-lived GW signal with strain amplitude%
\footnote{For simplicity, we model the GW as a monochromatic plane wave
$h^{\mu\nu} \approx h^{\mu\nu}_0 e^{i(\omega_g t - \vec{k}\cdot\vec{x})}$,
with ${\mu,\nu} = 0,1,2,3$ and $h^{\mu\nu}_0 \sim \mathcal{O}(h_0)$.}
$h_0 \sim 10^{-31}$ at Earth with $\omega_g \simeq 5~\mathrm{GHz}$. At MHz frequencies, corresponding to $M_{\rm bh} \sim 10^{-6} M_\odot$, the expected strain can be as large as $h_0 \sim 10^{-24}$~\cite{Franciolini:2022htd}. Recent work indicates that inspiral signals and eccentric orbits can further enhance the prospects, potentially generating additional harmonics in the HFGW band~\cite{Schenk:2025ria,Blas:2026ybh}. 

HFGWs may also arise from ultralight bosonic dark matter~\cite{Ferreira:2020fam}. If bosons of mass $m_\varphi$ exist, spinning black holes of mass
$M_{\rm bh} \simeq M_\odot (10^{-11}\,\mathrm{eV}/m_\varphi)$
can develop a macroscopic boson cloud through superradiance~\cite{Arvanitaki:2009fg,Brito:2015oca}. The cloud emits nearly monochromatic gravitational waves through different processes~\cite{Arvanitaki:2010sy}. The most relevant for our work are those generated by boson annihilation processes, $\varphi + \varphi \to h$, at a frequency corresponding to twice the mass of the boson, 
\begin{equation}
\omega_g \equiv 2\pi f_g = 2 m_\varphi\approx 
3 \times 10^{4}\,\mathrm{Hz}
\left(\frac{m_\varphi}{10^{-11}\,\mathrm{eV}}\right),
\label{eq:ULDMfr}
\end{equation}
while other emission channels are possible, in particular if self-interactions of the bosonic field are included~\cite{Arvanitaki:2010sy,Baryakhtar:2020gao, Witte:2024drg}.  For \mbox{$m_\varphi \gg 10^{-11}\,\mathrm{eV}$}, the emitted GWs lie in the HFGW regime; however, efficient superradiance at these masses requires sub-solar-mass black holes, implying a primordial origin and a nontrivial dark sector.

If such superradiant systems exist in the Milky Way, one expects strains of order $h_0 \sim 10^{-25}$ $(10^{-28})$ at MHz (GHz) frequencies~\cite{Berlin:2023grv}. In contrast to PBH binaries, the GW frequency in this case is fixed by the boson mass, necessitating active frequency scanning in resonant searches. 
The duration of the emission phase is typically of order days to months for the gravitational coupling $\alpha \equiv G M_{\rm bh} m_\varphi  \sim 0.1$ at GHz frequencies, decreasing for higher frequencies.  For these reasons, and because early PBH inspirals naturally provide broadband and slowly evolving signals, we focus primarily on PBH sources in the context of this discussion of \GravNet.

Finally, HFGWs can also originate from processes in the primordial Universe. Inflation, first-order phase transitions, thermal fluctuations, topological defects, and related phenomena generically produce stochastic gravitational-wave backgrounds with spectra that can extend into the MHz--GHz range~\cite{Aggarwal:2025noe,Ringwald:2022xif,Ghiglieri:2020mhm,Vagnozzi:2022qmc}. A particularly robust example is the so-called cosmic gravitational-wave microwave background, which is an unavoidable prediction of the Standard Model supplemented by inflation~\cite{Aggarwal:2020olq,Ringwald:2022xif}. These signals form a persistent and isotropic background arriving from the earliest moments of the Universe. Several models predict characteristic strains of order $h_0 \sim 10^{-29}$ $(10^{-34})$ at MHz (GHz) frequencies~\cite{Aggarwal:2020olq}.

\subsection{Detection Methods for High-Frequency Gravitational Waves}

Detecting gravitational waves (GWs) at frequencies above $\mathcal{O}(10~\mathrm{kHz})$ is challenging with established GW technologies, both because of the detector response and because of readout/noise limitations. For laser interferometers, the response of long Fabry--P\'erot arm cavities becomes strongly frequency dependent at $f_g \equiv \omega_g/(2\pi) \gtrsim \mathcal{O}(10~\mathrm{kHz})$, as the GW period becomes comparable to (or shorter than) the photon storage time and light travel time in km-scale arms~\cite{Maggiore:2007ulw}. Recent analyses have emphasized that, in the shot-noise limited regime, interferometers can nonetheless exhibit narrow high-frequency sensitivity features at multiples of optical free spectral ranges, and that detector configurations can in some cases be tuned to scan such narrow bands~\cite{Schnabel:2024hem,Jungkind:2025oqm,Heisig:2025oim}.%
\footnote{For instance, Refs.~\cite{Schnabel:2024hem,Jungkind:2025oqm} discuss high-frequency response and sensitivity structures of operating interferometers and the possibility of accessing the kHz--tens-of-kHz range by adjusting optical parameters.}
At the same time, extending interferometric searches deep into the $\gtrsim 10~\mathrm{kHz}$ regime requires dedicated characterization of the readout chain and instrumental backgrounds at these frequencies, which is currently not standard in the large-scale GW-detector program. 

A second traditional approach is based on mechanical resonators (``Weber bars'' and their modern descendants)~\cite{Maggiore:2007ulw}. These devices are typically optimized for operation near their mechanical resonance frequencies (historically in the kHz band), and their sensitivity generally degrades at much higher frequencies. Nonetheless, resonant acoustic and phonon-trapping concepts have been proposed to target higher-frequency bands~\cite{Goryachev:2014yra}. 

Overall, HFGW detection remains comparatively less explored than the sub-kHz regime. Motivated by the potentially transformative implications of any detection in the MHz-GHz band, a growing body of work has recently surveyed and advanced a range of HFGW concepts~\cite{Aggarwal:2020olq,Aggarwal:2025noe}. This is also reflected by new proposals and re-analyses using existing precision experiments, including re-interpretations of axion-haloscope data as HFGW searches~\cite{Kim:2025izt}. Since gravitation is universal and cannot be shielded, the passage of a GW can in principle imprint itself on essentially any precise laboratory setup; this perspective underlies many of the emerging HFGW detection strategies.

In \GravNet, we focus on one of the most promising directions for realistic progress in the near term across a broad frequency range: electromagnetic (EM) cavities operated in the presence of strong background EM fields, and in particular strong magnetic fields. This choice is motivated by (i) the existence of technically mature cavity and readout technology developed for axion and axion-like particle searches, and (ii) the ability to realize large and stable background fields $\mathcal{O}(10\,\mathrm{T})$ in compact detector volumes, enabling competitive GW-to-EM transduction per detector size and resource footprint. 

\begin{figure}[t!]
  \begin{center}
    \includegraphics[width=0.8\textwidth]{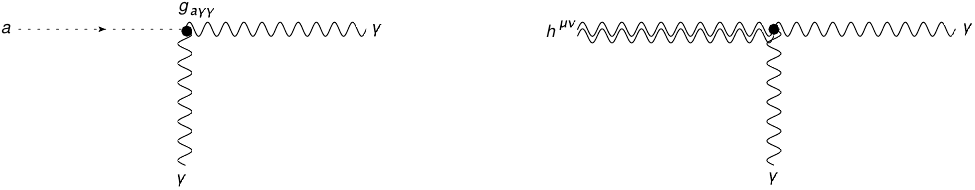}
  \end{center}
  \vspace{-10pt}
  \caption{Comparison of axion–photon conversion in a magnetic field (left) and gravitational-wave–to–photon conversion in a strong electromagnetic background (right), i.e. the quadratic interaction of EM fields $\gamma$ with GWs $h^{\mu\nu}$, illustrating the close analogy between the two processes exploited in cavity-based searches.}
  \label{fig:Feynm}
\end{figure}

Two main cavity-based mechanisms have been considered in recent years. The first is the modification of cavity eigenmodes due to effective boundary deformations induced by the GW (sometimes phrased as ``shape distortion'' or parametric modulation)~\cite{Ballantini:2005am,Berlin:2023grv}. The second is the generation of an effective EM current in the presence of a background EM configuration, commonly referred to as the inverse Gertsenshtein effect~\cite{Gertsenshtein:1962kfm,Berlin:2021txa}. The origin of the latter is straightforward: gravity couples universally to energy and momentum, and for EM fields the stress-energy tensor is quadratic in the field strength (or equivalently in $A^\mu$ at the level of interactions), implying an interaction of the schematic form shown in Figure~\ref{fig:Feynm}. As a consequence, in the presence of a background EM field $A^\mu_b$ and a GW $h^{\mu\nu}$, a signal field $A^\mu_s$ can be generated. For a background oscillating at frequency $\omega_b$, the induced signal appears at sideband frequencies $\omega_b \pm \omega_g$. In the following, we will mostly be concerned about the second mechanism, as it is the one where more progress can be achieved in the near future. Both are compared in~\cite{Berlin:2023grv}, and the inverse Gertsenshtein effect is the leading effect for frequencies close to the resonant frequencies of the cavities, which also supports focusing on this effect for a range of frequencies of the GWs.

The inverse Gertsenshtein effect is reminiscent of phenomenology from axion--photon interactions which, for an axion field $a$ with axion-photon coupling $g_{a\gamma}$, reads $\propto g_{a\gamma}a\,\vec{E}\cdot\vec{B}$, for which resonant radio-frequency cavities in strong magnetic fields constitute one of the leading experimental search strategies~\cite{Berlin:2022hfx}. A key implication is that there exists substantial theoretical and experimental expertise---including mature cryogenic cavity operation, low-noise RF amplification, and precision spectral analysis---that can be repurposed for HFGW detection. Indeed, recent work has demonstrated the feasibility of re-analyzing haloscope data to place competitive constraints on monochromatic HFGWs in the GHz band~\cite{Kim:2025izt}.

Two differences between cavity-based HFGW searches and cavity-based axion searches are worth emphasizing. First, many transient GW sources (e.g.\ compact-binary inspirals) naturally sweep through frequency, so the cavity does not necessarily need to be mechanically tunable, provided the signal remains within the resonant bandwidth long enough (e.g.\ $N_{\rm cyc}\gtrsim Q_l$ in the simplified discussion of Sec.~\ref{sec:sources}). In this case, the time evolution of $\omega_g(t)$ also sets the effective coherent integration time $t_{\rm int}$, since the signal eventually leaves the resonant band.

Second, unlike conventional axion dark-matter signals (which are expected to be quasi-stationary and approximately isotropic in the Galactic frame), GW signals have a definite propagation direction and polarization content, and they induce correlated responses across spatially separated detectors with predictable time delays. This enables additional discriminants and analysis handles that are central to \GravNet: coincidence tests, baseline-dependent correlation searches, and (for persistent signals) daily modulation and directional inference using Earth rotation.

\subsection{Interaction of High Frequency Gravitational Waves with Electromagnetic Cavities}
\label{sec:SignalPower}

In the following, we work explicitly in the framework of the inverse Gertsenshtein effect, where the interaction of a gravitational wave with a static magnetic field is described in terms of an effective electromagnetic current. In this context, for a cavity within a static magnetic field, $\omega_b=0$, the induced field $A^\mu_s$  in Figure~\ref{fig:Feynm}  oscillates at the frequency of the gravitational wave $\omega_g$. This mode can be resonantly enhanced using (microwave) cavities. The sensitivity to its detection depends on various factors, including  $\omega_g$, the incoming direction of the GW, the strength of the external magnetic field $B_0 = |\vec B_0|$, as well as the coupling of the incoming GW to the different resonance modes of the cavity. 

Assuming a GW signal described by a functional form of the strain $h(t)$, the inverse Gertsenshtein effect implies an induced effective current in an electromagnetic cavity in a static magnetic field, which is given by
\begin{equation}\label{eq:jeff}
j\simeq  B_0 \omega_g h_0 e^{i\omega_gt}.
\end{equation}
It is important to note that the above equation becomes more complicated for a realistic experiment, where the induced current depends on the orientation of the setup w.r.t. the incoming GW wave as well as its polarization, and the current $j$ has to be convolved with the resonance curve and therefore the quality factor $Q_l$ of the cavity. The width of the resonance curve is inversely proportional to the quality factor $Q$, and therefore large values of $Q$ yield a large resonant enhancement of the signal. 
This is particularly true if the signal lasts for enough cycles to fully `load' the cavity ($N_\mathrm{cyc} \gtrsim Q_{l}$) and reach a stationary regime. 
This has been used to build sensitivity estimates for searches of, for instance, PBHs~\cite{Gatti:2024mde}, since the signal frequency during the inspiral phase of the merger varies as 
\begin{equation}
\dot{f} = 4.62\times 10^{2} \left( \frac{M_{\text{PBH}}}{10^{-9}M_\odot}\right)^{5/3} \left(\frac{f_g}{\text{GHz}} \right)^{11/3} \frac{\text{GHz}}{\text{s}}\,,
\label{eqn:fdot}
\end{equation}
which yields values of $10^{-6}$\,GHz/s to $10^{9}$~GHz/s for PBH merger masses between $M_\text{PBH} = 10^{-15}M_\odot$ and \mbox{$M_\text{PBH} = 10^{-7}M_\odot$} for frequencies in the range of interest, $2\,\mathrm{GHz} \leq f_g \leq 8\,\mathrm{GHz}$. 
For comparison, the bandwidth of an electromagnetic cavity at a center frequency of 5\,GHz with quality factors of $25\times 10^3$ and $5\times 10^5$, corresponds to $2\times 10^{-4}$\,GHz and $10^{-5}$\,GHz. 
This leads to signal times within the cavity bandwidth between 10\,s, which is long enough to integrate the signal with existing analysis approaches, to as low as $10^{-14}$\,s, as can be seen in Tab.\,\ref{tab:TimeInBW}. 


\begin{table}[ht]
\renewcommand{\arraystretch}{1.0}
\setlength{\tabcolsep}{4pt}
    \centering
    \begin{tabular*}{\textwidth}{@{\extracolsep{\fill}}ccccccc}
    \toprule\midrule
    $M_{\rm PBH} [M_\odot]$ & $f_{\rm ISCO}$ [GHz] & $f_{0}$ [GHz] & $\dot{f}$ [Hz/s] & $\Delta t$ [s], $Q_l=5 \times 10^4$ & $\Delta t$ [s], $Q_l=1 \times 10^5$ & $\Delta t$ [s], $Q_l=5 \times 10^5$ \\ 
    \midrule
$10^{-15}$ & $2 \times 10^{9}$ & 2 & $5.87 \times 10^{2}$ & $6.82 \times 10^{1}$ & $3.41 \times 10^{1}$ & 6.82 \\
$10^{-15}$ & $2 \times 10^{9}$ & 5 & $1.69 \times 10^{4}$ & 5.92 & 2.96 & $5.92 \times 10^{-1}$ \\
$10^{-15}$ & $2 \times 10^{9}$ & 8 & $9.46 \times 10^{4}$ & 1.69 & $8.46 \times 10^{-1}$ & $1.69 \times 10^{-1}$ \\
\midrule[0.25pt]
$10^{-13}$ & $2 \times 10^{7}$ & 2 & $1.26 \times 10^{6}$ & $3.16 \times 10^{-2}$ & $1.58 \times 10^{-2}$ & $3.16 \times 10^{-3}$ \\
$10^{-13}$ & $2 \times 10^{7}$ & 5 & $3.64 \times 10^{7}$ & $2.75 \times 10^{-3}$ & $1.37 \times 10^{-3}$ & $2.75 \times 10^{-4}$ \\
$10^{-13}$ & $2 \times 10^{7}$ & 8 & $2.04 \times 10^{8}$ & $7.85 \times 10^{-4}$ & $3.92 \times 10^{-4}$ & $7.85 \times 10^{-5}$ \\
\midrule[0.25pt]
$10^{-11}$ & $2 \times 10^{5}$ & 2 & $2.72 \times 10^{9}$ & $1.47 \times 10^{-5}$ & $7.34 \times 10^{-6}$ & $1.47 \times 10^{-6}$ \\
$10^{-11}$ & $2 \times 10^{5}$ & 5 & $7.84 \times 10^{10}$ & $1.28 \times 10^{-6}$ & $6.38 \times 10^{-7}$ & $1.28 \times 10^{-7}$ \\
$10^{-11}$ & $2 \times 10^{5}$ & 8 & $4.39 \times 10^{11}$ & $3.64 \times 10^{-7}$ & $1.82 \times 10^{-7}$ & $3.64 \times 10^{-8}$ \\
\midrule[0.25pt]
$10^{-9}$ & $2 \times 10^{3}$ & 2 & $5.87 \times 10^{12}$ & $6.82 \times 10^{-9}$ & $3.41 \times 10^{-9}$ & $6.82 \times 10^{-10}$ \\
$10^{-9}$ & $2 \times 10^{3}$ & 5 & $1.69 \times 10^{14}$ & $5.92 \times 10^{-10}$ & $2.96 \times 10^{-10}$ & $5.92 \times 10^{-11}$ \\
$10^{-9}$ & $2 \times 10^{3}$ & 8 & $9.46 \times 10^{14}$ & $1.69 \times 10^{-10}$ & $8.46 \times 10^{-11}$ & $1.69 \times 10^{-11}$ \\
\midrule[0.25pt]
$10^{-7}$ & $2 \times 10^{1}$ & 2 & $1.26 \times 10^{16}$ & $3.16 \times 10^{-12}$ & $1.58 \times 10^{-12}$ & $3.16 \times 10^{-13}$ \\
$10^{-7}$ & $2 \times 10^{1}$ & 5 & $3.64 \times 10^{17}$ & $2.75 \times 10^{-13}$ & $1.37 \times 10^{-13}$ & $2.75 \times 10^{-14}$ \\
$10^{-7}$ & $2 \times 10^{1}$ & 8 & $2.04 \times 10^{18}$ & $7.85 \times 10^{-14}$ & $3.92 \times 10^{-14}$ & $7.85 \times 10^{-15}$ \\  \midrule\bottomrule
    \end{tabular*}
    \caption{Time GWs from a PBH merger stay within the bandwidth of a cavity $\Delta t$ depending on the cavities resonance frequency $f_0$, PBH merger mass $M_{\rm PBH}$ and loaded quality factor of the cavity $Q_l$. }
    \label{tab:TimeInBW}
\end{table}

\begin{figure}[ht]
  \begin{center}
    \includegraphics[width=\textwidth]{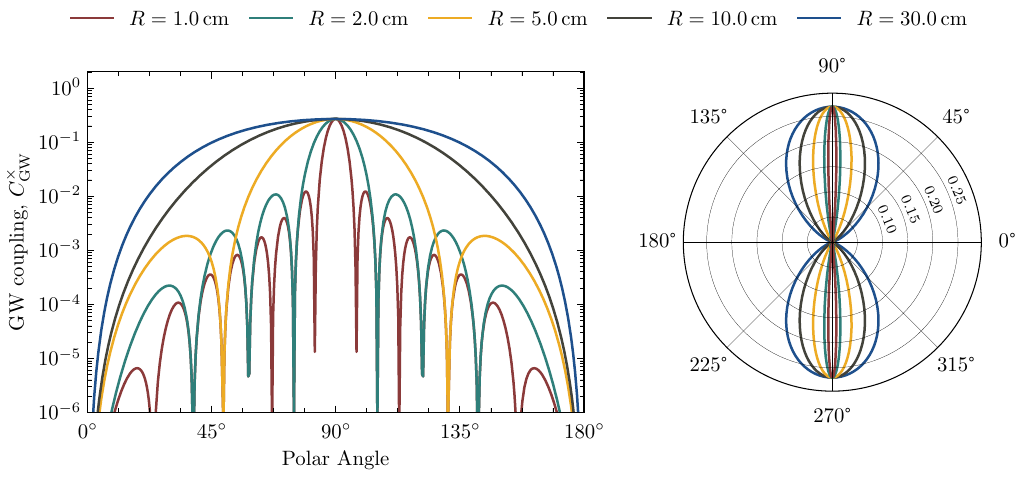}
  \end{center}
  \caption{The gravitational coupling $C_{\rm GW}^\times$ for an ideal TM$_{010}$ mode in a cylindrical cavity with fixed height $H=20$\,cm and various radii $R$. A uniform magnetic field is applied along the cavity symmetry axis. The coupling is maximized when the GW propagation direction is perpendicular to the magnetic field (at polar angles $90,\ 270^\circ$).
  }
  \label{fig:pureTM010_coupling}
\end{figure}

In order to estimate the experimental sensitivity, the deposited energy from HFGWs must be determined, which is given by the integrated deposited power. By approximating the resonance curve as a rectangular step function, where the width is given by the bandwidth $b=(f_0/Q_l)$, and assuming $P_\mathrm{sig}(\omega_g) = P_\mathrm{sig}(\omega_0)$ and $\dot{f}$ to be constant within the narrow bandwidth of the cavity, the deposited energy is calculated as~\cite{Gatti:2024mde}
\begin{equation}
E = P_{\rm sig}\frac{f_0}{Q_l} \frac{1}{\dot{f}}\,.
\label{eqn:energy}
\end{equation}
%
The EM power induced by the GW is calculated in the stationary limit where the cavity is fully rung up as~\cite{Kim:2025izt}
\begin{equation}
    P_{\text{sig}} = \frac{\beta}{1+\beta}\frac{\omega_{g}}{2\mu_{0}} h_{\times,+}^{2} \langle B_{0}^{2} \rangle Q_{l} C_{\rm GW}^{\times,+} V\,,
    \label{eq:TTpower}
\end{equation}
where $\beta$ is the antenna coupling coefficient, $V$ the volume of the cavity, $B_0$ the magnetic field strength, and $h$ the GW strain. The factor $C^{\times,+}_{\rm GW} = (\eta_{\rm GW}^{\times,+})^2 $ describes the coupling of the GW to the cavity mode as defined in~\cite{Kim:2025izt}, where $\eta_{\rm GW}$ is instead used in the conventions of~\cite{Berlin:2021txa,Schenk:2025ria}.
This coupling depends on the angle between the incoming GW and the $B$-field, ranging from $0$ to $10^{-2}$, where the maximum value and the average over all incident angles depend on the cavity mode and cavity geometry. This is illustrated for an idealized cylindrical cavity assuming the TM$_{010}$ mode in Figure \ref{fig:pureTM010_coupling} and typical values are given in Tab. \ref{tab:cylindrical_tm010_freq}.

A detailed calculation of the signal power is derived in the proper detector frame in~\cite{Berlin:2021txa} and in the transverse-traceless (TT) frame in~\cite{Kim:2025izt}. This calculation requires the choice of a reference frame (or `gauge') and some debate has been devoted to this choice in recent literature. Ultimately, the result cannot depend on the choice of frame as long as all signal contributions are accounted for. 
In proper detector coordinates, the background magnetic field is perturbed as calculated in~\cite{Berlin:2021txa}; however, the cavity walls are also displaced by the GW, leading to an additional signal. 
In this work, we assume that the GW frequencies are large enough that the elastic forces in the cavity walls are negligible~\cite{Schenk:2025ria,Gue:2026kga}, such that they behave as if they were freely falling. In this case, it is more convenient to use TT coordinates, in which the GW does not displace the cavity walls and the signal in Eq.~\eqref{eq:TTpower} only emerges through the Gertsenshtein current, Eq.~\eqref{eq:jeff}~\cite{Schenk:2025ria,Ratzinger:2024spd,Kim:2025izt,Gue:2026kga}


\begin{table}[ht]
\renewcommand{\arraystretch}{1.0}
\setlength{\tabcolsep}{4pt}
    \centering
    \begin{tabular*}{\textwidth}{@{\extracolsep{\fill}}ccccc}
    \toprule\midrule
    $M_{\rm PBH} [M_\odot]$ & $f_{g}$ [GHz] & $h_0$ & $h_c$ & $P_\mathrm{sig}$ [W] ($h = h_0$)  \\ 
    \midrule
$10^{-15}$ & 2.0 & $3.2 \times 10^{-30}$ & $2.7 \times 10^{-22}$ & $1.1 \times 10^{-41}$ \\
$10^{-15}$ & 5.0 & $6.0 \times 10^{-30}$ & $2.3 \times 10^{-22}$ & $9.6 \times 10^{-41}$ \\
$10^{-15}$ & 8.0 & $8.1 \times 10^{-30}$ & $2.1 \times 10^{-22}$ & $2.9 \times 10^{-40}$ \\
\midrule[0.25pt]
$10^{-11}$ & 2.0 & $1.5 \times 10^{-23}$ & $5.7 \times 10^{-19}$ & $3.6 \times 10^{-29}$ \\
$10^{-11}$ & 5.0 & $2.8 \times 10^{-23}$ & $4.9 \times 10^{-19}$ & $6.6 \times 10^{-29}$ \\
$10^{-11}$ & 8.0 & $3.8 \times 10^{-23}$ & $4.6 \times 10^{-19}$ & $9.0 \times 10^{-29}$ \\
\midrule[0.25pt]
$10^{-7}$ & 2.0 & $7.0 \times 10^{-17}$ & $1.2 \times 10^{-15}$ & $1.7 \times 10^{-22}$ \\
$10^{-7}$ & 5.0 & $1.3 \times 10^{-16}$ & $1.1 \times 10^{-15}$ & $3.1 \times 10^{-22}$ \\
$10^{-7}$ & 8.0 & $1.8 \times 10^{-16}$ & $9.8 \times 10^{-16}$ & $4.2 \times 10^{-22}$ \\
\midrule
\bottomrule 
    \end{tabular*}
    \caption{Electromagnetic power $P_\mathrm{sig}$ deposited in a cavity by a gravitational wave from a PBH merger of mass $M_\mathrm{PBH}$. To calculate the strain $h_0$ using Eq.~\eqref{eq:h0_pbh} at the resonance frequency $f_0 = f_g$, we consider a distance of one astronomical unit (AU) ( $\sim4.8\times 10^{-9}$ kpc). To calculate the deposited power, we use Eq.~\eqref{eq:TTpower} assuming the values discussed in Sec. \ref{sec:experimentalSetup} of $B_0 = 12\,\text{T}$, $C_\mathrm{GW} = 0.1$, $Q_l=Q_{\text{eff}}=\min(10^5, N_{\textrm{cyc}})$, and that the volume is independent of the frequency $V=3\times10^{-4}\,\text{m}^3$. }
    \label{tab:RFPower}
\end{table}

\begin{table}[htbp]
\renewcommand{\arraystretch}{1.0}
\setlength{\tabcolsep}{4pt}
\centering
\begin{tabular*}{\textwidth}{@{\extracolsep{\fill}}cccccc}
\toprule\midrule
Radius [cm] & Cylinder height [cm] & Volume [m$^3$]& TM$_{010}$ Freq [GHz] & Peak coupling $C^\times_\mathrm{max}$& Average coupling $C^{\times}_{\rm GW}$\\
\midrule
30 & 20 & $5.66\times 10^{-2}$ &  0.38 & 0.27& 0.16 \\
10 & 20 & $6.28\times 10^{-3}$ &  1.15 & 0.27 & 0.12\\
5  & 20 & $1.57\times 10^{-3}$ &  2.29 & 0.27 & 0.08\\
2  & 20 & $2.50\times 10^{-4}$ &  5.74 & 0.27 & 0.03\\
1  & 20 & $6.00\times 10^{-5}$ &  11.47 & 0.27 & 0.02\\
\midrule\bottomrule
\end{tabular*}
\caption{Resonance frequency of the TM$_{010}$ mode of a cylindrical cavity as a function of the cavity radius. Also shown are the assumed cavity heights and the average coupling factor to the cross polarization of a gravitational wave arriving from a random direction. Listed frequencies are given in GHz and rounded to two decimal places.}
\label{tab:cylindrical_tm010_freq}
\end{table}

Table \ref{tab:RFPower} gives an overview of the generated power in the envisioned cavities for various choices of the PBH mass under the assumptions described above. 
From Eq.\,\eqref{eqn:fdot},  $\dot{f} \propto f_g^{11/3}$ which, from \eqref{eqn:energy}, leads to a dependence of the deposited energy on the frequency of the GW $E\propto f_g^{-8/3}b$, where $b$ is the bandwidth of the resonator. 
For a PBH binary of equal masses $M_{\text{PBH}}$, the strain $h_0$ and characteristic strain $h_c$ are calculated following~\cite{Franciolini:2022htd} as
 \begin{equation}\label{eq:h0_pbh}
     h_0 = 9.77\times10^ {-34} \, \left(\frac{f_g}{\mbox{GHz}} \right)^{2/3} \, \left(\frac{M_{\text{PBH}}}{10^{-12}M_\odot }\right)^{5/3} \, \left(\frac{\mbox{kpc}}{D} \right),
 \end{equation}
 and
  \begin{equation}
h_c = 4.54\times10^ {-28} \, \left( \frac{\mbox{GHz}}{f_g} \right)^{1/6} \, \left( \frac{M_{\text{PBH}}}{10^{-12}M_\odot} \right)^{5/6} \, \left(\frac{\mbox{kpc}}{D} \right),
 \end{equation}
 where $D$ is the distance in kpc.
 As shown in~\cite{Schenk:2025ria,Blas:2026ybh,Borysenkova:2026tbd}, for a signal with time-evolving frequency, the amount of power that ends up in the different modes requires an analysis in the time domain yielding larger values compared to those considering the stationary limit we applied above. 
Furthermore, the presence of eccentricity is also a source of higher tones, which also injects more energy in a situation with narrow resonances.

In the following chapter, we discuss several experimental considerations on the choice of the magnet systems, the cavity designs as well as the readout.

\section{Experimental Setup}\label{sec:experimentalSetup}

\subsection{Magnet Systems}

In the first version of \GravNet, we foresee using two different types of magnet systems: one magnet system that can host a large-volume cavity with a base resonance frequency around 150\,MHz, and a second magnet system aimed at a significantly smaller cavity volume with resonance frequencies in the GHz regime. 

\begin{figure}[ht]
  \begin{center}
    \includegraphics[width=0.6\textwidth]
    {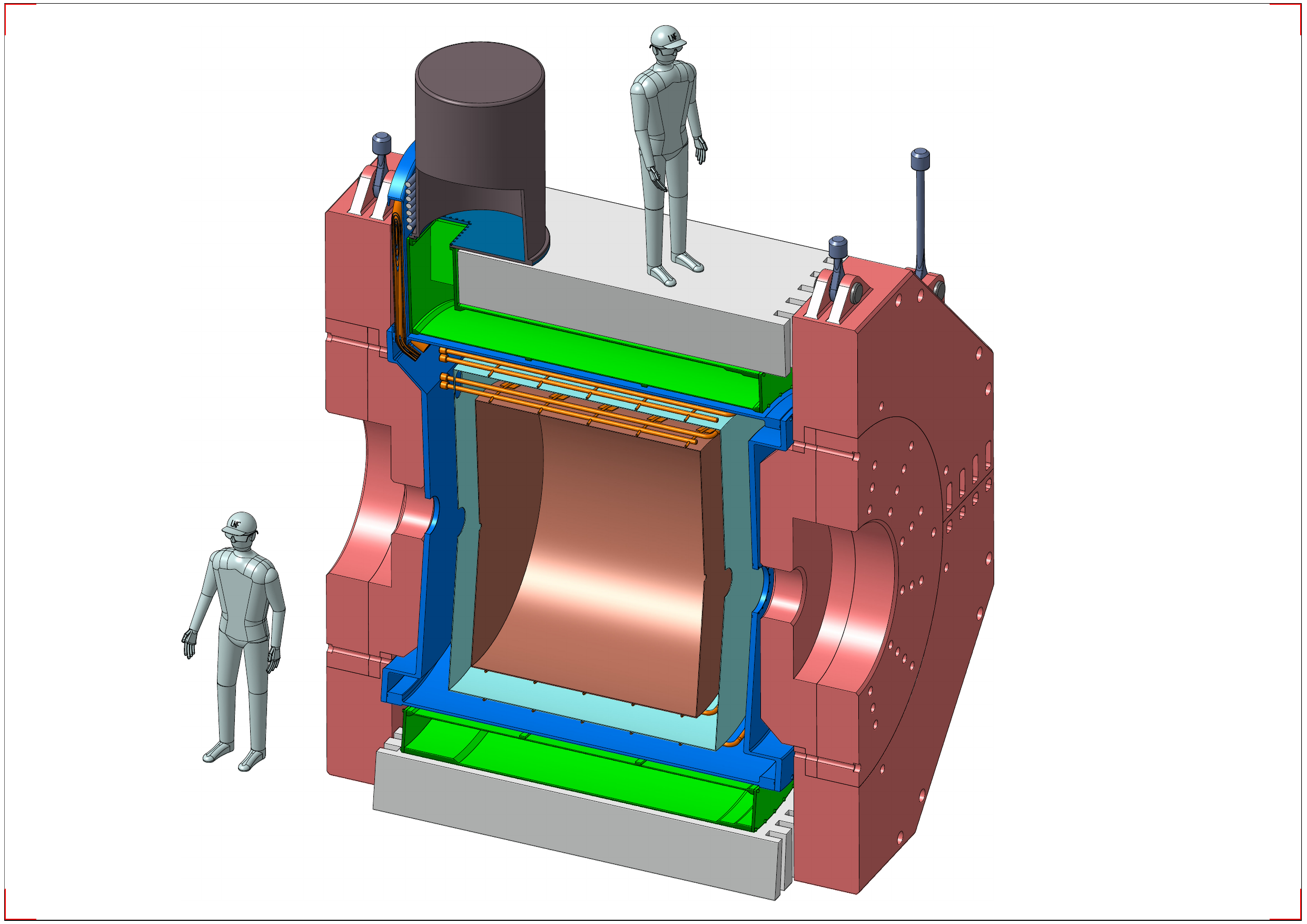}
    \includegraphics[width=0.35\textwidth]{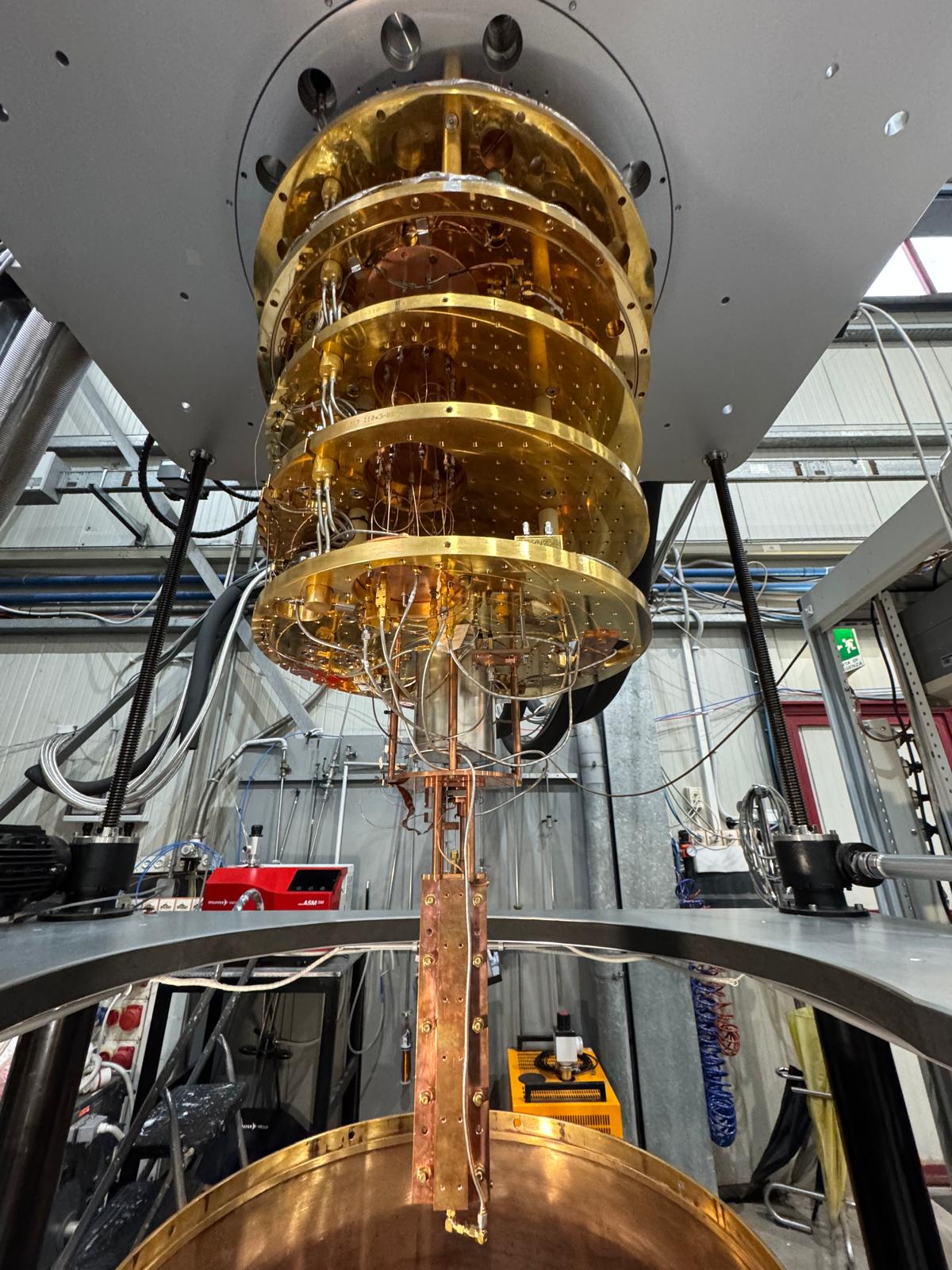}
  \end{center}
  \caption{Magnet and cryostat systems used in the \GravNet program. Left: Schematic cross section of the FLASH magnet, indicating the magnet bore, cryostat structure, cooling connections and vacuum system (credits Cesidio Capoccia). Right: BlueFors dilution refrigerator with the installed 9 GHz cavity system at LNF, illustrating the compact cavity assembly and associated cabling inside the cryostat.\label{fig:magnets}
  }
\end{figure}

The large volume cavity will be placed in the FINUDA magnet~\cite{bertani1999finuda, Modena:1997tz} at the Laboratori Nazionali di Frascati (LNF), which has an iron-shielded solenoid coil with a radius of 1.4\,m and a length of 2.2\,m; it is made from an aluminum-stabilized niobium titanium superconductor (Figure \ref{fig:magnets}). The magnet provides a homogeneous axial field of strength up to $B_0$ = 1.1\,T. This large volume makes it possible to install a large cavity with a fundamental mode at a relatively low resonance frequency of 150\,MHz. A newly constructed cryostat system will allow for the operation of the cavity at a temperature of 1.9\,K. We refer to this as the FLASH cavity~\cite{Alesini:2023qed} in the following. The relatively low magnetic field strength is compensated for by its large volume. Moreover, for PBH-merger signals at these `low' frequencies, the cavity ringing time reaches tens of milliseconds, exceeding the cavity lifetime and increasing the sensitivity as compared to mergers at higher frequencies. 

The second type of magnets will be commercial systems integrated in cryostat systems. We foresee the initial version of \GravNet having at least three such systems with a central magnetic field between 9 and 12\,T, allowing the operation of a cavity at a base temperature of 10\,mK (Figure \ref{fig:magnets}). The field average over the usable cylindrical volume of $R=5$\,cm and $H=30$\,cm is about 20\,\% lower than the central value.

\subsection{Cavity Design and Construction}
\label{sec:cavityDesign}

The shape of the cavities defines their base resonance frequencies, their higher resonance modes, and also their coupling to HFGWs, labeled as $C_{\rm GW}^{\times,+}$ in  Eq.\,\eqref{eq:TTpower}. Moreover, a trade-off between the volume of the cavities and the available volume of the magnetic field must be made. 
A significant advantage of cavities for \GravNet compared to cavities in the context of axion-search experiments is that they do not need to be tunable to be sensitive to a significant parameter space of transient signals. This reduces the required R\&D effort tremendously compared to dedicated axion-search experiments such as ADMX~\cite{ADMX:2010ubl}. Given that the construction processes for large cavities with volumes of several liters are significantly different from those for smaller cavities with a volume of several centiliters, these processes are discussed separately in the following sections. 

For the baseline calculations of sensitivities within \GravNet, we assume one cylindrical cavity placed in the FLASH magnet system optimizing the available volume. To achieve the lowest possible frequency with our largest magnet system, thus maximizing the signal a GW will induce, we foresee a cylindrical resonant cavity made of copper with a radius of 980\,mm and a length of 1600\,mm, corresponding to an inner volume of approximately 4.80\,m$^3$~\cite{Alesini:2023qed} (Figure~\ref{fig:magnets}).
This optimizes the use of the available volume within the magnet and yields a resonance frequency of the $\rm{TM}_{010}$ mode of 129\,MHz. The quality factor is expected to be about $Q\sim5\times10^5$~\cite{Alesini:2023qed}.  The cavity will be hosted in a dedicated custom cryostat and cooled to 2\,K by a cold compressor using LHe) from the LNF cryogenic plant. The cryostat and cryogenic components will be designed at LNF and purchased and installed within the \GravNet project. We foresee that the cryostat will be composed of an external stainless-steel vacuum vessel, containing an aluminum-alloy radiation shield surrounding the cavity kept at about 70\,K by cold gaseous helium (GHe). Both the cavity and the shield will be cooled in contact with pipes with flowing helium. The cryogenic plant~\cite{ligi:2002} is already available at LNF.

\begin{figure}[ht]
  \begin{center}
    \includegraphics{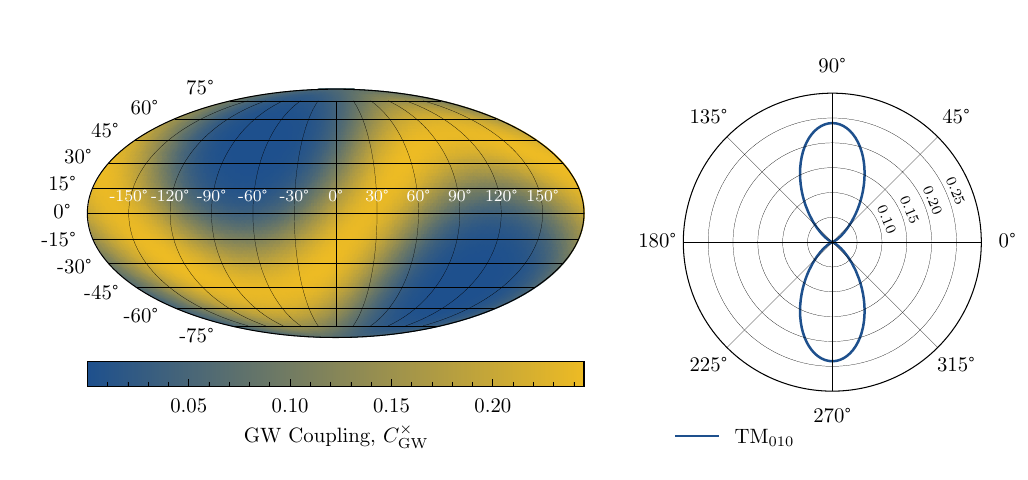}
  \end{center}
  \caption{\textbf{Left:} The GW coupling $C_\mathrm{GW}^\times$ for different sky positions of the GW source. The detector is located at LNF at 0:00 AM, and the magnetic field is aligned with the local zenith.
  \textbf{Right:} The GW coupling of a cross-polarized gravitational wave to the TM$_{010}$ mode of the FLASH cavity. 
 \label{fig:FLASH_coupling_TM010}
  }
\end{figure}

The gravitational wave coupling to the TM$_{010}$ mode was numerically evaluated for the given geometry of the FLASH cavity. In this mode, the electric field is dominated by the axial component $E{z}$, which provides the main contribution to the GW cross-polarization coupling, $C^{\times}_\mathrm{GW}$. Figure~\ref{fig:FLASH_coupling_TM010} presents $C^{\times}_\mathrm{GW}$ as a function of the GW incident angle measured with respect to the magnetic-field direction, $\hat{z}$. The coupling exhibits a clear angular dependence and reaches a maximum value of $C^{\times}_{\max}=0.24$ at a particular incident angle. This behavior indicates that the coupling efficiency is governed by the interplay between the TM$_{010}$ field geometry and the GW polarization and incidence configuration.
To incorporate the directional dependence under realistic observing conditions, we also evaluated the coupling for GWs arriving from various sky directions for the LNF site at 0:00 AM, as summarized in Figure~\ref{fig:FLASH_coupling_TM010}. This calculation accounts for the location- and time-dependent transformation between the sky coordinates and the cavity (laboratory) frame. The resulting distribution demonstrates that the effective coupling of the FLASH cavity TM$_{010}$ mode depends on the GW propagation direction, implying that the observation time and/or the instrument orientation can modulate the achievable coupling (and hence the signal sensitivity).


The situation is different for commercial magnet systems, as their cylindrical volume typically has a significantly larger height compared to their diameter, for example, $R=4.5$\,cm and $H=27$\,cm are typical parameters. 
In order to find the optimal layout of cavities given this volume, several aspects must be considered. For now, we discuss only two cases. The first is a cylindrical cavity that fills the full volume. In this case, the angular average of the coupling coefficient to a cross-polarized gravitational wave for the dominant TM$_{010}$ mode (operating at $2.55$\,GHz) is $C^\times_{\mathrm{cyl}} \approx 0.05$.
The second case is that of $N$ spherical cavities in the same magnetic volume, operated simultaneously, where at most $N=H/(2R)$, cavities can be placed. With $R=4.5$\,cm, there are three degenerate dominant modes TM$_{m11}$ at frequency $2.91$\,GHz for $m\in \{-1, 0, 1 \}$, with a combined average coupling coefficient of $C^\times_{\rm{sph}} \approx 0.20$. As detailed in Sec.\,\ref{sec:networkOperation}, the signal-to-noise ratio (SNR) for GW signals scales as the number $N$ of simultaneously operated and coherently read out cavities. Comparing the two cases,

\[r \equiv \frac{\mathrm{SNR}_{\rm{cyl}}}{\mathrm{SNR}_{\mathrm{sph}}} \approx \left( \frac{\pi R ^2 H}{4/3 \pi R^3 } \right) \cdot \frac{1}{N}  \cdot \frac{C^\times_{\mathrm{cyl}}}{C^\times_{\mathrm{sph}}}   = \left( \frac{H}{4/3 R} \right) \cdot \frac{2R}{H}  \cdot  \frac{C^\times_{\mathrm{cyl}}}{C^\times_{\mathrm{sph}}} \approx \frac{3}{2} \cdot \frac{C^\times_{\mathrm{cyl}}}{C^\times_{\mathrm{sph}}} \approx 0.38, \]
however the coupling for cylindrical cavities is the lowest for large $H/(2R)$, as seen in Table~\ref{tab:cylindrical_tm010_freq}. Instead, taking three cylindrical cavities with height $H=9.0\,$cm and radius $R=4.5\,$cm, for which $C_\mathrm{cyl}^\times \approx 0.12$, reduces the SNR ratio to only $r=0.90$, so, generally, to maximize the average coupling to the dominant mode, it is best to avoid high $H/(2R)$ cylindrical cavities. Hence, it is advantageous to operate multiple cavities with an aspect ratio closer to unity compared to one elongated cavity. However, coherently adding the contributions from multiple cavities presents an additional experimental challenge.

We recognize that the SNR also depends on the operating frequency $f$ and the quality factor $Q_0$ of the cavity.
Typically for a cylindrical cavity, the frequency will scale with height $H$, however, it is height independent for the mode TM$_{010}$. Because of this, and a similarity in frequency between the two compared cavity types, we only consider purely geometric factors for this estimate. 
A simulation comparing the intrinsic quality factor of a spherical and cylindrical cavities with radius $R=4.3\,$cm and height of $H=9.0\,$cm reveals a 27\,\% higher $Q_0$ of the spherical cavity compared to the cylindrical one, assuming the same conductivity of $4\times10^9$\,S/m corresponding to OFHC copper. 
This increases the SNR ratio in favor of spherical cavities to $r = 0.71$.

As discussed in Sec.~\ref{sec:SignalPower}, the response of a resonant cavity to HFGWs depends on the cavity quality factor $Q$ as well as all the different modes that may be excited and the time dependence of the signal. 
Increasing $Q$ simultaneously enhances the on-resonance signal power and narrows the resonance bandwidth, thus increasing the number of scan steps needed to cover a certain frequency interval. For a stationary monochromatic signal, assuming a fixed measurement time, this leads to a linear dependence of the SNR with the loaded cavity quality factor $Q_{l}$ in case $Q_l \ll Q_s$, where $Q_s$ is the effective quality factor of the GW signal. In this regime, the cavity bandwidth limits the measurement.
If instead $Q_l \gg Q_s$, the SNR scales proportionally to $\sqrt{Q_l}$ because the signal power grows $\propto Q_l$ while the integrated noise decreases as $1/\sqrt{Q_l}$~\cite{Chaudhuri:2019ntz}.
In the special case $Q_l \sim Q_s$, one obtains the scaling ${\rm SNR}\propto Q_l\sqrt{Q_s}\sim Q^{3/2}$.
For transient HFGW signals considered with durations shorter than or comparable to the cavity ring-up time, the relevant observable is the total energy deposited in the cavity rather than the steady-state power. 
The deposited signal energy integrated over the signal duration is, to a good approximation, independent of $Q_l$ as can be seen in equation \eqref{eqn:energy}, assuming that the readout time matches the signal duration given by $t_s = \Delta \nu / \dot{f} = f_0/(Q_l\dot{f})$. 
Considering the time integral over the noise power as $N=2\Delta\nu t$ independent measurements of band-limited white noise the variance of the accumulated energy becomes $\mathrm{Var}(E) = 2\left< E \right>^2 / N = (k_BT)^2\Delta\nu t$.
Hence, the SNR scales as $E/\sigma_E \propto Q$.
If the cavity fully rings up and the signal duration is longer than the readout (integration) time, the steady state case is recovered. In any case, a large unloaded quality factor $Q_0$ and therefore low losses in the cavity itself are always beneficial. 

\begin{figure}[ht]
  \begin{center}
    \includegraphics[width=0.48\textwidth]{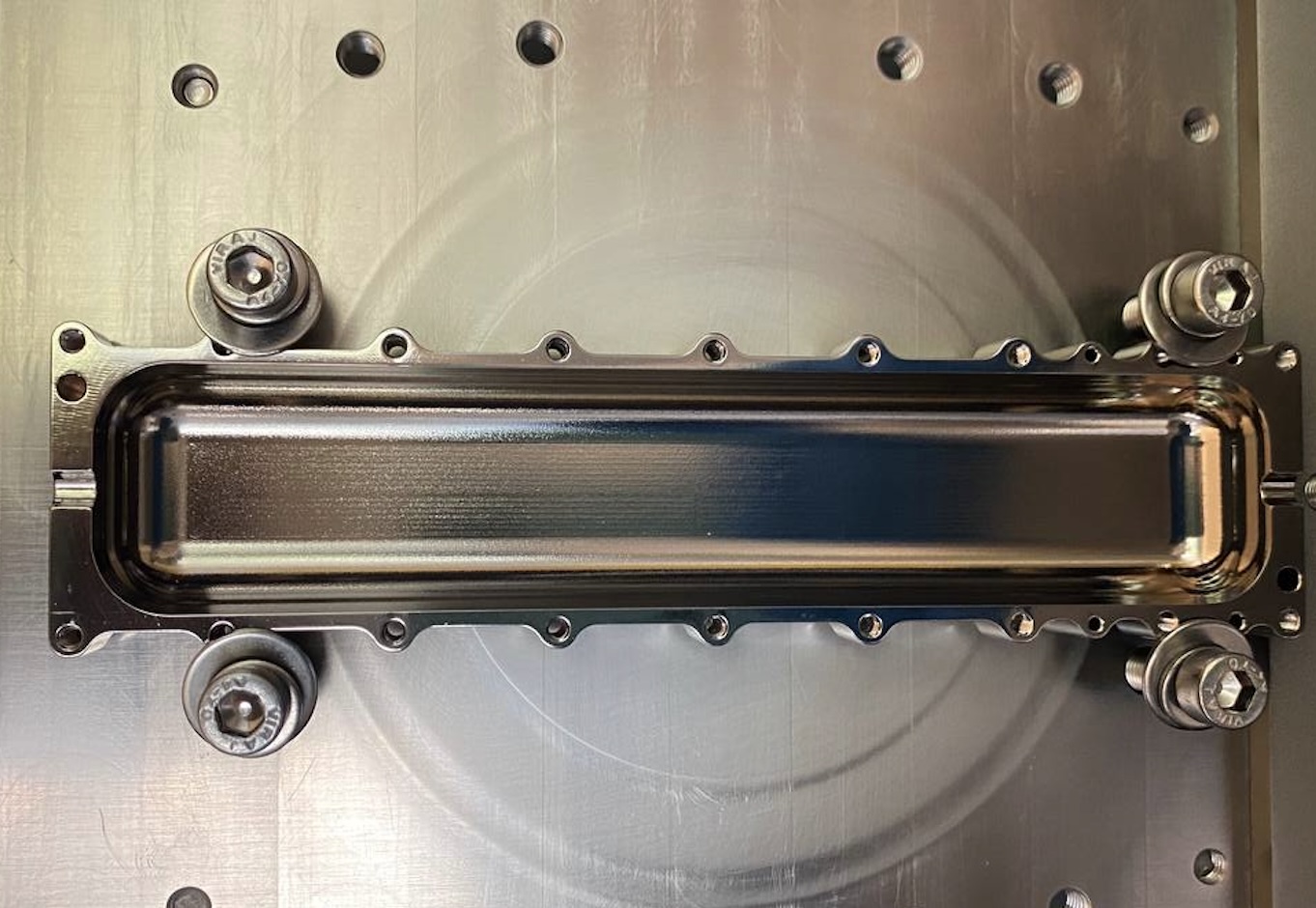}
    \includegraphics[width=0.5\textwidth]{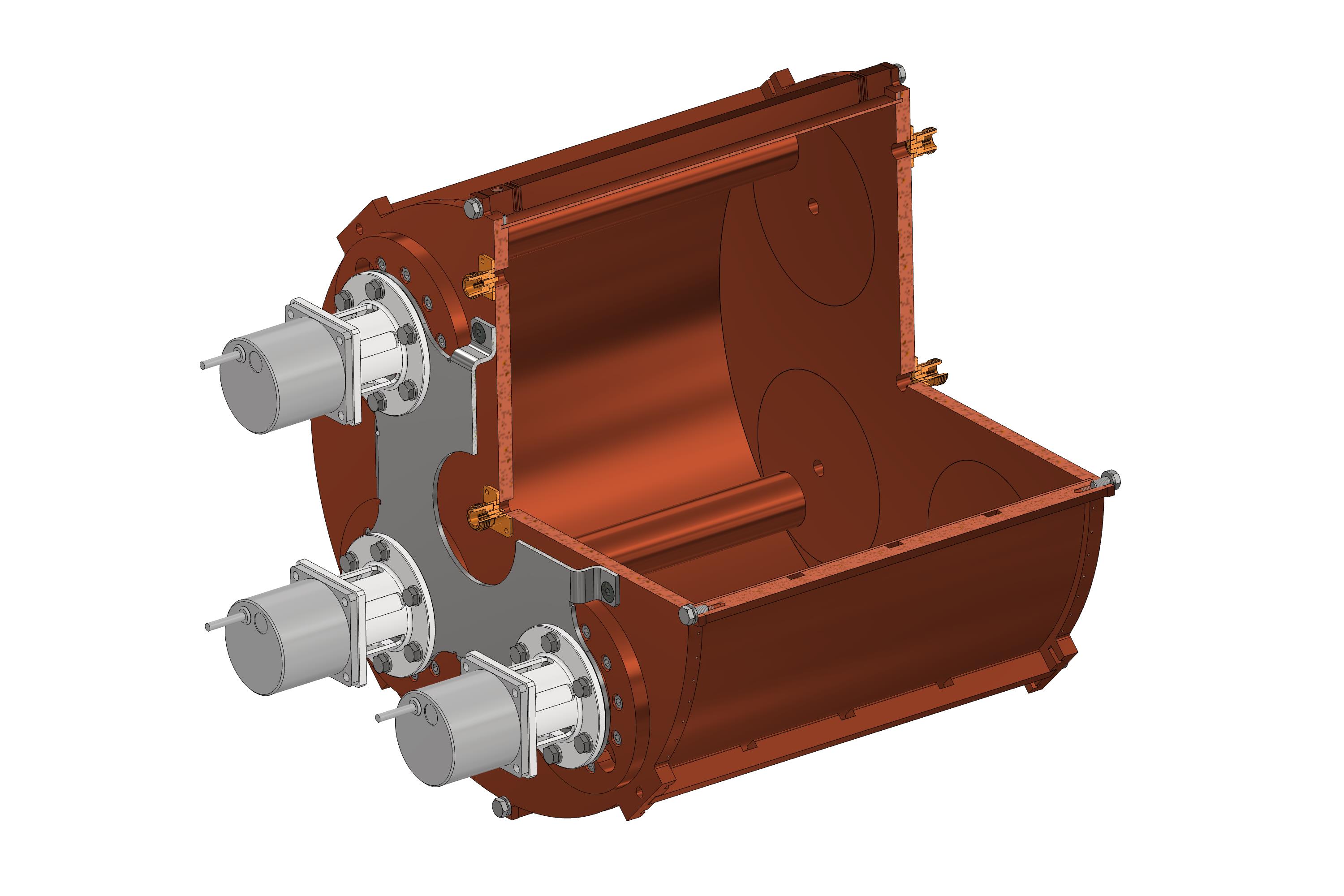}
  \end{center}
  \caption{Left: NbN coated cavity studied in the Supax setup in~\cite{Schmieden:2024wqp}. Right: schematic drawing of the FLASH cavity prototype ($1/6$ scale compared to the original).  \label{fig:cavities}
  }
\end{figure}

Simple high-frequency cavities can be milled into a block of solid copper, yielding unloaded $Q$-factors on the order of $5-10\times10^4$~\cite{Schneemann:2023bqc}. There are two promising avenues for enhancing the quality factor of a cavity: coating the inner surface with a superconductor is the first approach, with an expected enhancement of the quality factor by a factor 10 to 1000. Characterizations of niobium–titanium, niobium–tin and rare-earth barium copper oxide (ReBCO) cavities in a strong magnetic field have already been conducted~\cite{AlimentiSCCavity,QUAXNBTI}. Moreover, first successfully operated ultra-high $Q$ cavities using ReBCO in strong magnetic fields have been already demonstrated at the Center for Axion and Precision Physics Research (CAPP)~\cite{CAPPCavity,PhysRevApplied.17.L061005}. Moreover, tests using niobium nitride coating (Figure\,\ref{fig:cavities}) were also reported in ~\cite{Schmieden:2024wqp}. 
The second approach targets the use of dielectric cavities, which have shown an enhancement of $Q$ by a factor of $10^3$ over superconducting cavities in strong magnetic fields, at the cost of a smaller effective volume for the cavity mode~\cite{Alesini:2022lnp,CAPPCavity}. 
While ReBCO and dielectric cavities can already be used for HFGWs searches, they apply only to cylindrical cavities, due to the nature of these materials. Fabrication of cavities with more complex geometries, such as spherical cavities, requires a different deposition technique, such as sputtering. Besides, the position of the antennas within the cavities must be optimized. Overall, it seems that quality factors on the order of $5\times 10^5$ are reachable. 

Another interesting venue for cavity optimization is the use of metamaterials to artificially change the plasma frequency of the cavity which modifies the relation between the geometry of the cavity and its resonance frequencies. This can be achieved, for example, by filling the cavity volume with a regular grid of wires. Changing the distances of the wires changes the resonance frequency while the effective volume remains constant. This is in particular interesting for higher frequencies, where usually the cavity volume becomes rather small. This approach, also known as {\it plasma haloscope} has been studied recently in~\cite{Lawson:2019brd,Capdevilla:2024cby,Lindahl:2025zyc, Kowitt:2023wob} and in particular in the context of GW searches in~\cite{Capdevilla:2024cby}.

\subsection{Basic Cavity Readout Systems}
\label{sec:readout}
The simplest readout scheme of a typical cavity-based axion-search experiment --- and for this matter also for HFGW experiments --- is based on a real-time spectrum analyzer that is capable of streaming in-phase and quadrature (IQ) time-series data to a readout computer. This approach is limited by intrinsic (e.g. thermal and electronic) noise, however, is simple to realize and builds therefore the baseline to which we compare the advanced readout schemes in the following. 

The optimal linear amplifiers are based on quantum technologies. Their intrinsic noise is due to quantum fluctuations of the signal whose minimum value at the input of the amplifier is bounded to twice the zero-point energy for the mode $\omega$ of interest, $\hbar \omega/2$, and is usually expressed as an effective temperature $T_{\mathrm{SQL}}=\hbar \omega/k_B$. 
The contribution to the noise power is then $P_{\mathrm{SQL}}=\hbar \omega\sqrt{\Delta\nu/t}$~\cite{PhysRevD.26.1817, PhysRevD.88.035020}, where $\Delta\nu$ is the signal bandwidth which corresponds to the cavity bandwidth $b$ for transient signals and $t$ the acquisition time. Different types of quantum amplifiers are used for axion experiments as well as superconducting-qubit readout~\cite{Gatti:2021sar}. It is also a well-established technology that is commercially available. Depending on the frequency range and bandwidth, three different types of amplifiers can be distinguished: Microstrip Superconducting Quantum Interference Device (MS SQUID), Josephson Parametric Amplifiers (JPAs), and Superconducting Travelling Wave Amplifiers (TWPAs)~\cite{Muck_2010,Zhong,Roch}. We foresee using a SQUID for the readout of the FLASH cavity and TWPA and JPA amplifiers for the small GHz cavities. For the latter, we also investigate the single-photon-detection approach. 

\begin{figure}[t]
  \begin{center}
\includegraphics{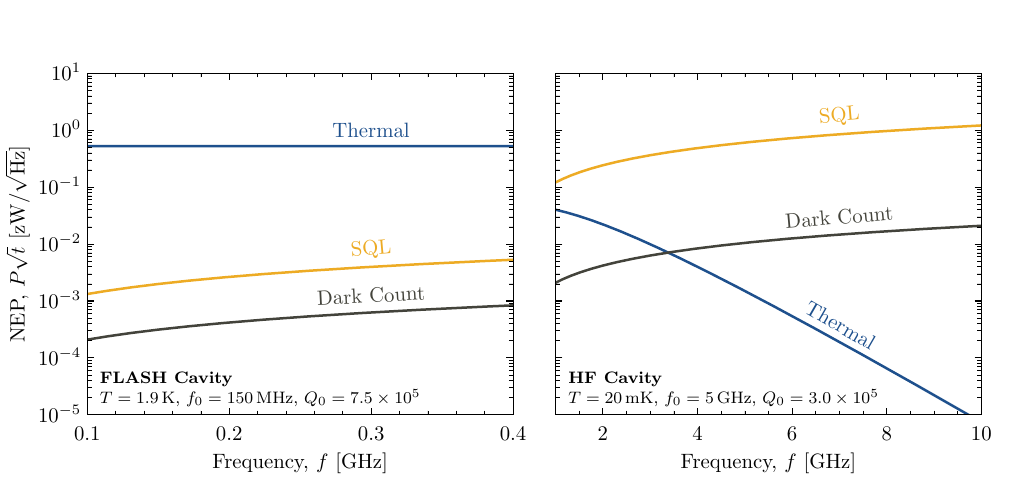}
  \end{center}
  \caption{The noise equivalent power (NEP) with frequency $f$ for the dominant noise contributions: thermal noise, dark count (using a dark count rate of $10\,\mathrm{Hz}$), and quantum noise at the standard quantum limit (SQL). These are computed by dividing the relevant noise power by the frequency resolution, equivalent to $P \sqrt{t}$, where $t$ is the total integration time. \textbf{Left:} The noise spectral density for the FLASH cavity system. \textbf{Right:}  The noise spectral density for the high-frequency (HF) cavity system. Both of these configurations are summarised in Tab.\,\ref{tab:NoisePower}.}
  %
  \label{fig:noise}
\end{figure}

An overview of expected noise power for the different readout technologies vs. the cavity resonance frequency is shown in Figure\,\ref{fig:noise} where: the r.m.s of the thermal power from a resonator of quality factor $Q_l$ is 
\begin{equation}
\label{eq:thermalnoise}
\sigma_{\overline{P}_{\mathrm{th}}}=\frac{\sigma_{E_{\mathrm{tot}}}}{t}=hf\sqrt{n(T)(n(T)+1)}\sqrt{\frac{f}{{Q_l t}}}\,,
\end{equation} 
where $h$ is the Planck constant, $f$ is the photon frequency, and $n(T)$ is the photon occupation number in the cavity at temperature $T$. The standard quantum limit (SQL) of an amplifier, that includes amplifier added noise and vacuum fluctuations, is $P_{\rm SQL}= hf \sqrt{f/Q_l t}$; the noise $P_{\rm dark \, counts}= hf \sqrt{\nu_{\rm DC}/t}$ where $\nu_{\rm DC}$ is the device dark count rate~\cite{KuzminCBJJ}.

\subsection{Multimode Readout Development for the MHz Cavities}

HFGWs couple to TM and TE modes with similar strength. Depending on their polarization, we will acquire both types of modes from the FLASH cavity in the frequency range between 100 and 300\,MHz. 
Two distinct antennas will couple to TM and TE modes, a dipole and a loop antenna, respectively. A Microstrip SQUID Amplifier (MSA)~\cite{Muck:1999ts, Muck_2010} kept at 1.9\,K is expected to be an optimal solution with a noise added by the amplifier between 200 and 400~mK, in terms of low noise, frequency band and gain, for the first stage of signal amplification. As second amplification stage a cryogenic heterojunction field-effect transistor (HFET) amplifier will be employed. The signal will be split at room temperature and bandpass filters will select the modes of interest before digitization. Multichannel fast ADCs  together with FPGA based signal processing will be used to acquire the signals. The offline combination of five modes with a 2\,$\sigma$ threshold will allow, for instance, a background reduction by 9 orders of magnitude, sufficient to detect transient events. Preliminary studies on this readout scheme have been already successfully conducted ~\cite{Alesini:2023qed}. In summary, we expect a noise level of $T_{\rm sys} = 2.2\,$K in the final readout system.

\subsection{Multimode Readout Development for GHz Cavities}

Microstrip SQUID amplifiers are well suited for readout at frequencies up to approximately 1\,GHz. In this regime, their noise performance is dominated by dissipation in the shunt resistors required for stable SQUID operation, which prevents them from reaching the quantum limit but still allows for near–quantum-limited amplification with excellent linearity and dynamic range~\cite{Muck_2010}. However, at higher resonance frequencies, the noise performance of SQUID-based amplifiers degrades, and alternative technologies are required.

For cavities operating at several GHz, Josephson parametric amplifiers (JPAs) and travelling-wave parametric amplifiers (TWPAs) constitute promising alternatives. JPAs~\cite{Zhong} can operate at the quantum limit and are frequency tunable, but they typically provide only narrow instantaneous bandwidths (of order tens of MHz), making them less suitable for simultaneous readout of multiple cavity modes. In contrast, TWPAs~\cite{Roch} operate efficiently at frequencies above approximately 4\,GHz and offer significantly larger bandwidths in the GHz regime, ranging from about 1\,GHz for four-wave-mixing designs up to roughly 4\,GHz for three-wave-mixing implementations. This wide bandwidth makes TWPAs particularly attractive for multimode or multiplexed readout schemes relevant for \GravNet. Their main drawback is a higher added noise compared to JPAs, typically corresponding to a noise temperature of a few $\hbar\omega/k_B$.

In practice, several readout architectures are conceivable. Signals from different cavity resonances may be frequency-multiplexed and amplified by multiple JPAs, or alternatively amplified simultaneously using a single broadband TWPA. While the detailed design of the readout chain remains to be optimized, initial studies—based, for example, on Ref.~\cite{Aybas:2021nvn}—indicate that a total system noise temperature of $T_{\mathrm{sys}} \simeq 300 \,\mathrm{mK}$ is achievable for GHz-frequency cavities.


 \subsection{Single-Photon-Counting Readout}\label{sec:SPD}

At frequencies above a few GHz the quantum noise $\hbar\omega$, rising linearly with frequency, spoils the benefits of working with a dilution refrigerator. To get around this limit, a considerable effort is ongoing to exploit quantum techniques such as squeezing, quantum metrology, and quantum sensing to achieve optimal measurement strategies for axion-dark-matter searches~\cite{Brady,Maccone,Shi2023}. It is therefore the aim to develop dedicated single-photon quantum sensors with a detection efficiency higher than 50\% for photons with frequency between 1 and 10\,GHz, with a dark-count rate of less than a few Hz, so that the contribution given to noise is less than the thermal one.  

Single-photon detection at microwave frequencies with circuits based on superconducting transmon qubits~\cite{Koch} was achieved in 2007~\cite{Schuster}. We already successfully performed qubit spectroscopy and quantum sensing experiments~\cite{app14041478,11249713,Piedju,DElia:2025ztz,Supergalax,DEliaCBJJ}.
Of interest for cavity experiments operating at large magnetic field is the itinerant photon detection, that allows keeping the superconducting electronics away from the $\vec B$-field region. Several experimental schemes for itinerant photons, based on superconducting qubits, have been proposed and realized~\cite{Inomata, Lescanne, Besse, kono, Supergalax,DEliaCBJJ}, but still with dark-count rates larger than 100\,Hz. Alternatively, the development of a magnetic field resistant qubit~\cite{DElia:2025ztz} would be a game changer enabling the possibility to adopt quantum non demolition (QND) schemes as well as searches through direct qubit excitation~\cite{PhysRevD.110.115021}. QND measurements enable repeated measurement of the presence of a single photon, opening up the way to error corrections and to drastic reduction in dark-count rates. QND measurements have been successfully used for the detection of photons generated inside a cavity~\cite{Dixit}. Within \GravNet, we propose an extension to two qubits of the scheme in~\cite{kono} , where a controlled Z-gate~\footnote{A change of sign of the qubit state $|1\rangle$ controlled by the arrival of a single photon}  between an incoming itinerant photon and a transmon qubit mounted inside a resonator was used to detect the photon arrival and reflection~\cite{Piedju}. In this new scheme, the dark-count rate scales with $p(1|0)^2$ instead of linearly, where $p(1|0)$ is the probability of erroneously measuring the qubit state 1 when it is 0. As  $p(1|0)$ can be as low as 0.1\%, this corresponds to a reduction in dark counts by up to three orders of magnitude. The minimum sensitivity reachable by this scheme for the amplitude of a coherent signal generated for a time duration $t_{\rm signal}$ is on the order of
\begin{equation}
    \overline{n}_{\rm min}=1/N_{\rm measure}=\tau_{\rm measure}/t_{\rm signal},
\end{equation}
where $N_{\rm measure}$ is the number of repeated measurements in a time $t_{\rm signal}$ and $\tau_{\rm measure}$ is the single measurement duration. In this scheme $\tau_{\rm measure}\sim 1/\gamma_{\rm cavity}$, where $\gamma_{\rm cavity}$ is the bandwidth of the conversion cavity, so that the integrated probability to produce a photon is 
\begin{equation}
    \int \dot{n}\,{\rm d}t\sim\gamma_{\rm cavity}\,\overline{n}\, t_{\rm signal}\sim 1.
\end{equation}
Given the quasi non-demolition nature of the measurement, probing successively and repeatedly with $N_q$ qubit-sensors the same coherent state would lead a factor $\int \dot{n}{\rm \, d}t\sim 1/N_q$ below the single photon sensitivity.
Entanglement of two or more qubits, such as the Greenberger--Horne--Zeilinger $|GHZ\rangle=(|000\dots 000\rangle+|111\dots 111\rangle$ state~\cite{RevModPhys.89.035002}, will be considered to further improve the sensitivity, although it is known that these states are very weak and subject to decoherence. Further enhancement of the sensitivity to strain, can be obtained by injecting squeezed-vacuum states into the resonant cavity as in~\cite{Backes}. The ultimate sensitivity might be obtained by combining squeezing and photodetection as discussed in~\cite{Maccone,Shi2023}. 


It is illustrative to discuss the single photon counting with a concrete example. For a cavity with a resonance frequency of 5\,GHz, the corresponding photon energy is $E_\gamma \approx 3.3\times10^{-24}\,\textrm{J}\,=\, 2\times10^{-5}\,$eV. In a naive picture, this is the minimal energy that must be deposited in the cavity to create at least one photon. However, photon creation is a quantum mechanical process, where the likelihood for a single-photon creation is following a Poisson distribution with mean $\lambda = E_{\rm HFGW} / E_\gamma$, where $E_{\rm HFGW}$ is the deposited energy of the HFGW. Hence, even tiny deposited energies that are smaller than the single photon energy could yield real photons, however, linearly suppressed with $\lambda$.

\section{Network Operation}
\label{sec:networkOperation}

The numerous advantages of detector networks in the context of fundamental physics have been convincingly demonstrated not only for gravitational-wave detection (LIGO--Virgo--Kagra~\cite{KAGRA:2013rdx}), but also for various families of dark-matter detectors, starting from the global network of optical magnetometers for exotic physics searches (GNOME~\cite{Afach2021_GNOME_NP,Afach2023WhatCanGNOME}), atomic-clock networks such as GPS.DM~\cite{Roberts:2017hla}, as well as currently emerging multi-sensor networks~\cite{Afach2023WhatCanGNOME} based on diverse sensors such as atomic spectrometers designed to detect fast (up to 100\,MHz) oscillations of fundamental constants~\cite{Tretiak2022_RFDMPRL} and ferromagnetic detectors of galactic axion-like particles (SHAFT~\cite{Gramolin2021search}).

For the initial version of \GravNet, we foresee operation with a large cavity at the FLASH magnet as well as nine small-scale cavities at the laboratories in Bonn, Frascati and Mainz, where three cavities are operated together in identical setups at each site. In the following discussion, we consider combining the signals of the nine GHz cavities, operating at similar frequencies around 6\,GHz. Two further simplifications are made: first, we ignore the fact that higher modes can be additionally read out for each cavity, which would yield additional information and hence increase the sensitivity. Second, we ignore the timing difference of incoming signals at the three network nodes, which depends on the location of the HFGW source and can range between 0 and $\approx$3\,ms.

Operating multiple HFGW detectors as a network provides a robust way to enhance sensitivity, reject instrumental backgrounds, and extract signal coherence. We discuss how the signal-to-noise ratio (SNR) scales with the number of detectors $N_\mathrm{cav}$, distinguishing between two detection strategies: continuous power readout and single-photon counting. The scaling depends critically on whether the signal is combined coherently or incoherently and on the statistical nature of the dominant noise.
In this work, we distinguish between the intrinsic SNR used in spectral analyses from correlation detection strategies. For a network of $N_\mathrm{cav}$ detectors linked incoherently, the intrinsic SNR scales at most as $\sqrt{N_\mathrm{cav}}$, while in the ideal case of a coherent combination it scales linearly with $N_\mathrm{cav}$. The detector network provides further advantages from correlation strategies, in which each detector scans a time window $\Delta T$ for discrete candidates. In this case, increasing the number of detectors leads to a strong suppression of the false alarm probability for a fixed intrinsic SNR.

In a power readout scheme the measured output of the $i$-th detector may be written as
$P_i(t) = P_{\rm sig}(t) + n_i(t)$,
where $P_{\rm sig}(t)$ denotes the signal contribution and $n_i(t)$ represents stochastic noise, assumed to be uncorrelated between different detectors.
If the detector outputs are combined incoherently at the power level, the total signal scales linearly with the number of detectors, while the noise adds in quadrature. The resulting network SNR therefore scales as
${\rm SNR}_{\rm net} = \sqrt{N_\mathrm{cav}}\cdot {\rm SNR}_{\rm single}$.
If, on the other hand, the signal phase information is preserved and the detector outputs can be combined coherently, the signal amplitudes add linearly while the noise remains incoherent. In this optimal case, the network SNR scales as
${\rm SNR}_{\rm net} = N_\mathrm{cav}\cdot {\rm SNR}_{\rm single}$.
Achieving this scaling requires precise relative timing and phase calibration between detectors, as well as a sufficient signal coherence time.


The situation changes when considering a single-photon readout in each cavity. Here, the crucial aspect is the
dark-count rate, or background rate of single-photon detection per time step $\Delta t$. For a single detector, the expected number of signal and background counts are denoted by $\mu_{\rm sig}$ and $\mu_{\mathrm{bkg}}$, respectively, and are assumed to follow Poisson statistics.
For a network of $N_\mathrm{cav}$ independent detectors, the total expected signal count scales as
\begin{equation}
\mu_{\rm sig}^{\rm net} = N_\mathrm{cav} \mu_{\rm sig},
\end{equation}
while the statistical uncertainty is dominated by background fluctuations,
\begin{equation}
\sigma_{\rm net} = \sqrt{N_\mathrm{cav} \mu_{\mathrm{bkg}}}.
\end{equation}
The resulting network SNR is therefore given by
\begin{equation}
{\rm SNR}_{\rm net}
= \frac{N_\mathrm{cav} \mu_{\rm sig}}{\sqrt{N_\mathrm{cav} \mu_{\mathrm{bkg}}}}
= \sqrt{N_\mathrm{cav}}\,{\rm SNR}_{\rm single}.
\end{equation}
Because photon counting is intrinsically incoherent, this $\sqrt{N_\mathrm{cav}}$ scaling represents the optimal sensitivity improvement achievable by increasing the number of detectors.
For transient signals, additional discrimination against background can be achieved by requiring temporal coincidence between photon detection events in multiple detectors. Coincidence requirements significantly suppress false-positive rates by reducing the effective background, thereby enhancing detection confidence.

Consider a network of $N_\mathrm{cav}$ single-photon detectors, each operated in independent measurement windows of duration $\Delta t$. Let $p_{\rm sig}$ be the probability for a detector to register a signal photon if a GW is present, and $p_{\mathrm{bkg}}$ the probability for a background (noise) count, per measurement window. A $k$-fold coincidence is defined as the detection of photons in at least $k$ out of $N_\mathrm{cav}$ detectors within the same measurement window.
For arbitrary $p_{\rm sig}$, the probability for a $k$-fold coincidence in a single measurement window is given by the binomial sum
\begin{equation}\label{eq:binomSumSignal}
p_{\rm sig}^{(k)} = \sum_{j=k}^{N} \binom{N}{j} 
\left(p_{\rm sig}\right)^j \left(1-p_{\rm sig}\right)^{N-j},
\end{equation}
and similarly for accidental coincidences due to noise,
\begin{equation}
p_{\mathrm{bkg}}^{(k)} = \sum_{j=k}^{N} \binom{N}{j} 
\left(p_{\mathrm{bkg}}\right)^j \left(1-p_{\mathrm{bkg}}\right)^{N-j}.
\end{equation}
In the limit of rare events, $p_{\mathrm{bkg}} \ll 1$, these reduce to the familiar approximation
\begin{equation}
p_{\mathrm{bkg}}^{(k)} \simeq \binom{N}{k} (p_{\mathrm{bkg}})^k.
\end{equation}
Over $N_{\rm t}$ independent measurement windows, the expected number of background coincidences is $\mu_{\mathrm{bkg}} = N_{\rm t} \, p_{\mathrm{bkg}}^{(k)}$. With one GW event per year, the expected number of detected signal coincidences scales as $\mu_\mathrm{sig} = 1\cdot p_\mathrm{sig}^{(k)}$.

Assuming Poisson statistics for the background, the standard deviation of background coincidences is 
\begin{equation}
{\rm SNR}_{\rm net} = \frac{\mu_\mathrm{sig}}{\sqrt{\mu_\mathrm{bkg}}} = \frac{p_{\rm sig}^{(k)}}{\sqrt{N_\mathrm{t} \, p_{\mathrm{bkg}}^{(k)}}}.
\end{equation}
In the rare-event limit $p_{\mathrm{bkg}} \ll 1$ and $p_\mathrm{sig} \approx \mathcal{O}\left(1\right)$, one can approximate

\begin{equation}
{\rm SNR}_{\rm net} \simeq \frac{1} {\sqrt{N_\mathrm{t}{\binom{N_\mathrm{cav}}{k}}}} \frac{p_{\rm sig}^{(k)}}{p_{\mathrm{bkg}}^{k/2}}.
\end{equation}
This shows that increasing the number of detectors strongly enhances the SNR: for linearly increasing  coincidence order $k$, alongside $n$ (fixed $n-k$ term), 
as long as $p_{\mathrm{sig}}\gg p_{\mathrm{bkg}}$, we expect an exponential increase in the signal to noise ratio. For larger signal probabilities, the exact binomial sum must be used, and the SNR scaling with $N_\mathrm{cav}$ eventually saturates as the coincidence probability approaches unity.


In short, for both scenarios, the network operation will boost the statistical sensitivity and allows for a much better control of systematics and rejection of false signals. Examples are given in Sec. \ref{sec:singePhotonDetection}. For network synchronization, each experimental site will be equipped with a GPS-referenced data-acquisition system offering several channels of analog-to-digit converters (ADC), several digital channels, and GPS-based time synchronization at a level of better than 200\,ns. 

The key decision to be made (see below) is how much of the raw data needs to be stored and transferred versus `smart pre-processing' of the raw data (which could reduce the amount of `byte trafficking' by many orders of magnitude at the expense of limiting the opportunities for post-processing raw data). 

\subsection{Triggering and Data Transfer}

In the case of a single-photon readout, the expected data rate per cavity will be minimal and corresponds to the background or noise rate of the system. The situation is significantly different for a continuous readout scheme, where the voltage $V(t)$ needs to be recorded for each cavity. The expected raw data rate per cavity depends on the digitized bandwidth, precision of the digitized values, and the number of modes to be read out. Typically 14~bits per value are used and a bandwidth of several MHz, yielding data rates of tens to hundreds of MB/s. This would be necessary to both coherently add the signal across the entire network and thus optimize the SNR. The data volume per site that hosts three cavities could therefore accumulate to several tens of TB per day, requiring a stable high-speed connection to the other participating institutions. Hence a dedicated online analysis system has to be developed which analyses the recorded data instantaneously and only keeps such information, where local upwards fluctuations are present. The relevant time-steps are marked and sent to all other participating sites, so that the corresponding information is kept also at their locations for further analysis. We expect to reduce the necessary data-transfer across all sites by at least a factor of 20. In a common data-center, a combined analysis will then be performed on a daily basis, and only the potentially interesting data blocks will be retained. In addition, it is foreseen to implement a random trigger, which keeps around 2\% of the data on each site in addition to the signal-triggered events. With this approach, we expect to reduce the overall data volume stored per site and per day of data taking to at most 1\,TB. 

\section{Analysis Considerations}

A divers network of GW detectors offers a multitude of analysis possibilities, where the analysis approach depends not only on the targeted signal source (transient or monochromatic), but also on the detector technologies employed within the network which could resolve the time-structure of the signal, e.g. using a time-resolved power measurement or only detect the presence of a signal, and could be sensitive at similar frequencies or cover a larger span of frequencies. 

In the initial phase of \GravNet each detector is responsible for analysing its data stream in real time and flag time periods containing a potential signal. This 'trigger' signal is distributed among all active detectors in the network, which will then save the raw detector data within the time-span in question to facilitate a combined analysis at a later stage. 
The stored time-window takes into account the spatial distance of the detectors of 200 to several thousand kilometers, leading to significant shifts in arrival time of the signal of up to  40\,ms which are potentially longer than the duration of a transient signal.

The most simple combination of data from different network sites is a coincidence analysis, which only requires a yes/no decision 
from each detector and allows the combination of arbitrary detector technologies. The general principle is discussed in Sec. \ref{sec:networkOperation} and sensitivity estimates are given in Sec. \ref{Subsec: Combining sites}. 
For transient events, the direction of the source can be located by measuring the relative differences of the arrival time of the signal if at least four stations measure a signal.

If the time-evolution of the GW signal can be measured, more advanced combinations of the time-series data can be imagined, similar to what current GW observatories do. This would allow for a more precise determination of the source position, as well as deducing insights on the parameters of the signal source.  

Real-time analysis of detector data largely depends on the detector technology used. 
If the GW waveform is reconstructed, e.g. in the case of a classic power readout, the signal can be recovered from the detector and thermal noise by means of matched filtering or machine learning approximation of optimal filter. A yes/no decision on the presence of a signal could be taken e.g. utilizing modern methods of anomaly detection like variational autoencoders.
If multiple detectors, e.g. multiple cavities or the readout of multiple cavity modes from the same cavity, are used at one site, the signals could be added coherently if the frequencies are identical or combined in a search for transient signals locally to reduce the background rate. 

If more sensitive technologies like single photons counting or other quantum metrology methods are used the information on the waveform of the signal is typically lost. Those detectors require specific readout protocols that vary depending on the detector technology and are not discussed in detail here. The out is always a yes/no decision on a potential signal photon, and the state of many such detectors can easily be combined in a coincidence-type experiment, as mentioned above. 

In the next chapter, sensitivity prospects are presented for some individual detectors using RF cavities with classic power detection and single photon counting, as well as the impact of the combination of signals in a coincidence style experiment. 

\section{Performance of GravNet}

As an example case, we discuss the sensitivity of \GravNet to transient gravitational wave signals from PBH mergers with masses between $10^{-7}$ and $10^{-16}M_\odot$, where we estimate the maximal distance from Earth of those merger events that are still detectable with our terrestrial experiment. This can be converted to an upper limit on the density of primordial black holes in our galaxy, assuming, in addition, a certain length of data taking, which we take to be one year in the following. The preliminary estimates presented in the following should give the reader an idea of the achievable sensitivity.

In Sections \ref{Subsec:Expected Sens} and \ref{sec:singePhotonDetection} 
the sensitivities achievable with a single cavity are discussed using power readout with parametric amplification and advanced readout techniques employing single-photon counters, respectively. The combination of different detectors at different locations is discussed in Section \ref{Subsec: Combining sites}. 

%



\subsection{Expected sensitivities of a baseline system of GravNet}
\label{Subsec:Expected Sens}

In the baseline \GravNet system, we foresee operation of the sub-gigahertz FLASH Cavity at a temperature of 1.9\,K, while nine gigahertz-range cavities will be operated simultaneously at $T=$ 20\,mK. We expect achieving quality factors of $Q^{\mathrm{LF}}_0 = 7.5\times 10^5$ and $Q_0^{\mathrm{HF}} = 3\times 10^5$ in the large and small cavities, respectively. Targeting an antenna coupling factor of $\beta = 1$ when operating at fixed frequencies, the loaded quality factor will be $Q_l = 0.5\,Q_0$.

The readout system for all cavities in the baseline version of \GravNet will be realized with parametric amplifiers as the first stage, either JPAs or TWPAs, yielding, in both cases, a system noise temperature about $T_{\rm sys} = 300\,\text{mK}$ for the high-frequency setups. The low-frequency setup is dominated by the 1.9\,K thermal noise, which yields a system noise temperature about $2.2$\,K adding $0.3$\,K from the SQUID amplifier. 
The SNR is calculated using the approach to evaluating the RF power generated in the cavity discussed in Section \ref{sec:SignalPower}, in particular, Eq.\,\ref{eq:TTpower}. The noise power depending on the system noise temperature is calculated as
\begin{equation}\label{eqn:noisePower}
    P_{\text{noise}} = k_BT_{\text{sys}}\Delta\nu\,,
\end{equation}
where $\Delta\nu$ integrated bandwidth of the readout, taken to be 10\,kHz in the HF setup and 200\,Hz in the FLASH setup.
The resulting noise power is shown in Tab.\,\ref{tab:NoisePower}.
\begin{table}[ht]
\renewcommand{\arraystretch}{1.0}
\setlength{\tabcolsep}{4pt}
    \centering
    \begin{tabular*}{0.85\textwidth}{@{\extracolsep{\fill}}lccccc}
    \toprule\midrule
         Setup &  $f_0$ [GHz] & $Q_0$ & $T_{\text{sys}}$ [K] & $P_{\text{noise}}$ [W] & $h^{\textrm{min}}$\\
          \midrule
       HF &  5.0 & $3.0\times10^5$ & $0.3$ & $4\times 10^{-20}$ & $1.7\times 10^{-20}$\\
       FLASH & 0.15 & $7.5\times10^5$ & 2.2 & $6\times 10^{-21}$ & $2.2\times 10^{-21}$\\
       \midrule\bottomrule
    \end{tabular*}
    \caption{Noise power for different cavity setups and the corresponding strain sensitivity assuming SNR = 0.08 (see text). For the HF cavity the amplifier noise is the dominating noise source whereas for the FLASH cavity the 1.9\,K black-body radiation of the cavity itself is the dominant noise source. }
    \label{tab:NoisePower}
\end{table}

The minimally detectable signal is given by the SNR required for detection at a given confidence level.
Assuming a measurement time of one year with order $10^{10}$ measurement intervals and less than one false positive signal within the measurement period leads to a false positive rate (FPR) for an individual measurement of $<10^{-10}$ which corresponds to a threshold of $6.36\,\sigma$ above the noise. This translates into a required SNR of 8, if the confidence level of a positive signal is set to 95\% using $ \textrm{SNR} = \Theta + \Phi^{-1}(c_1)$, where $c_1$ is the desired confidence level, $\Phi^{-1}$ is the inverse of the cumulative distribution function of the normal distribution, and $\Theta$ is the threshold in sigmas of a normal distribution corresponding to the desired false positive rate.
Note that for different signal-duration assumptions, the readout window is adjusted accordingly to optimize sensitivity. This can quickly lead to tightened requirements on the FPR rate to $10^{-12}$, only slightly increasing the threshold to $7\, \sigma$ and hence the required SNR to 8.6.
The needed power for a positive identification of a signal is drastically reduced if the (known) shape of the signal in the time-domain is exploited by means of matched filtering. This enhances the SNR at a given signal power by up to a factor of 100 \cite{Sathyaprakash_2009} under ideal conditions.
Hence, the needed maximum signal power is taken to be $0.08\,P_{\text{noise}}$. The resulting minimally detectable strain $h^\mathrm{min}$ is calculated using Eq.\,\ref{eq:TTpower} and the cavity properties described in Sec.\,\ref{sec:cavityDesign} and presented in Tab.\,\ref{tab:NoisePower}.

\begin{figure}[ht]
    \centering
    \includegraphics{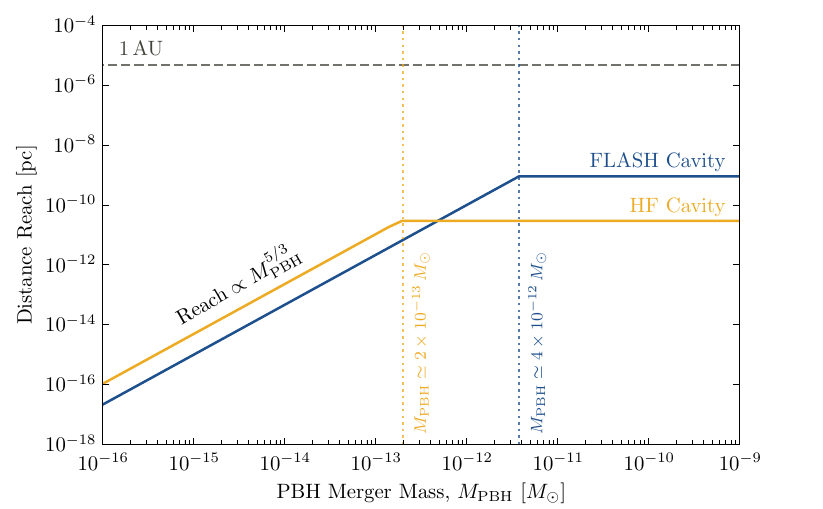}
    \caption{The distance reach for observing a PBH merger event at a given PBH merger mass $M_\mathrm{PBH}$ at a $95\%$ confidence level and a false positive rate below $10^{-10}$. This requires an SNR of at least $0.08$ assuming a linear readout of the cavity signals. Larger PBH masses lead to faster transient signals, which eventually cannot fully ring up the cavity. At this point, the distance reach becomes constant, occurring at the PBH merger masses indicated by the vertical lines. Also marked is $1\,\mathrm{AU} \simeq 4.8\times10^{-6}\,\mathrm{pc}$.}
    \label{fig:LinearAmp_DistanceReach}
\end{figure}

In this setup, the strain sensitivity is largely independent of the duration of the signal and hence the PBH mass as long as the cavity is fully rung up. The reason is that the the voltage in the cavity is proportional to the strain of the signal which scales with $M_{\rm PBH}^{5/3}$ while the time the signal spends within the bandwidth of the cavity scales inversely with $t\propto 1/\dot{f} \propto 1/M_{\rm PBH}^{5/3}$ and the SNR gain of the matched filter is proportional to the time of the signal \cite{1057571}. Hence, it is most instructive to show the reach in distance for a given PBH mass, which is displayed in Figure\,\ref{fig:LinearAmp_DistanceReach}. 
For very fast transient signals the cavity may not be fully rung up. This is modeled by using an effective quality factor for the cavity defined as $Q_{\rm eff} = \min(Q_l, N_{\rm cyc})$, where $N_{\rm cyc}$ is the number of cycles the signal spends within the bandwidth of the cavity. 
With only a few photons produced in the cavity, the power readout scheme fails and one must turn to advanced techniques like single photon detection. These are discussed in  Sec.\,\ref{sec:singePhotonDetection}.

Using a resonant structure in the experiment yields the best sensitivity close to the resonance frequency $f_0$, outside of which the resonant enhancement drops like $1/(f_0 - f)^2$. 
Even considering that the signal power induced by the GW rises proportional to the frequency of the gravitational wave $\omega_g$, see Eq.\,\ref{eq:TTpower}, 
the sensitivity drops quickly when moving away from the resonance frequency. 
In the intermediate frequency range, one typically finds a multitude of higher order resonant modes which can be exploited to achieve sensitivity at several frequencies simultaneously with a single cavity. The gain in sensitivity to transient signals will be studied in future work.

\subsection{Network operation}
\label{Subsec: Combining sites}

In this section, we illustrate the implications of the network formalism discussed above by considering concrete benchmark scenarios. We focus on transient HFGW signals and assume an expected event rate of order one detectable signal per year traversing the detector network located at Bonn, Mainz, and Frascati. Any impact due to the detector orientation is negligible, as can be seen from Fig. \ref{fig:pureTM010_coupling} as the angular separation is at most $9^{\circ}$.

We denote by $p_{\rm sig}$ the single-cavity detection efficiency, defined as the probability that a cavity registers a measurable response when an HFGW signal passes through the network. This efficiency is independent of the specific readout scheme (continuous power readout or single-photon counting), as it reflects the probability that a given signal strength produces a detectable response within a fixed time window. For the following examples, we assume a representative single-cavity detection efficiency of $p_{\rm sig} = 0.9$, i.e.\ a HFGW crossing the network is detected with probability 0.9 in each cavity.

The background probability per cavity and per measurement window is denoted by $p_{\mathrm{bkg}}$. We assume a measurement window of duration $\Delta t = 3~\mathrm{ms}$, corresponding to the characteristic time resolution discussed in Sec.~\ref{sec:networkOperation}. Over one year of operation, this yields $N_{\mathrm{t}} = \frac{365 \times 24 \times 3600{\rm\,s}}{\Delta t}\simeq 10^{10}$
independent measurement windows. In order to ensure robust detection, the expected number of accidental background coincidences over the full network and the full data-taking period must be well below unity. This motivates the use of coincidence measurements across multiple detectors.

\begin{figure}[t]
  \begin{center}
    \includegraphics[width=\textwidth]{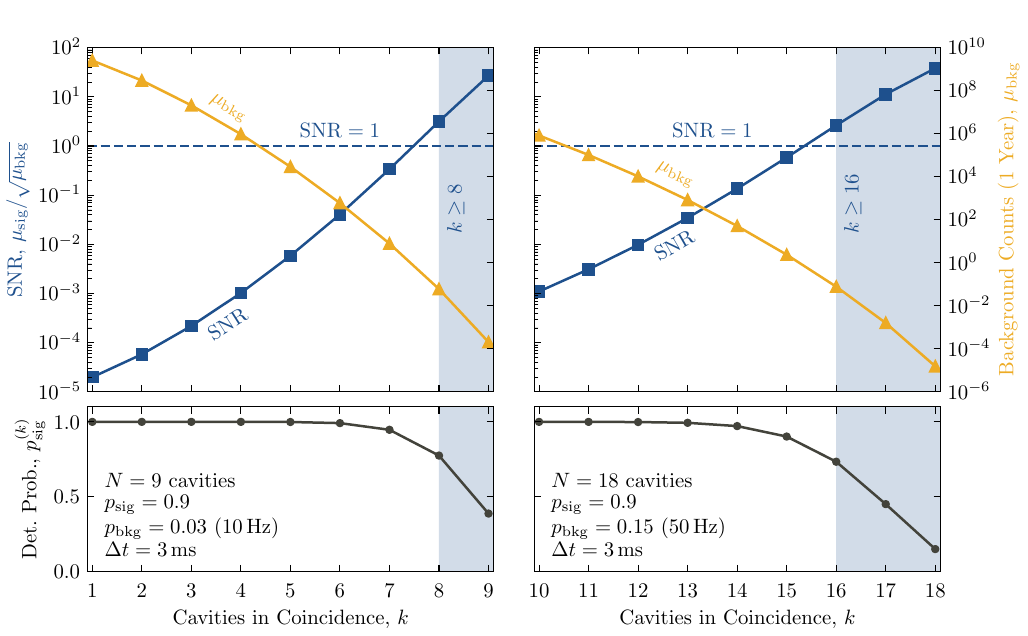}
  \end{center}
  \caption{Expected signal-to-noise ratio (SNR) and background performance over one year with required cavity coincidence order $k$ for a network of cavities. Also shown are the probabilities to detect a true HFGW signal for a given coincidence order, $p_\mathrm{sig}^{(k)}$. The assumed cavity detection efficiency is $p_{\rm sig}=0.9$ and the measurement window is taken to be $\Delta t = 3\,\mathrm{ms}$. \textbf{Left:} The performance of a network of $N=9$ cavities with assumed background probability per cavity and per measurement window of $p_{\mathrm{bkg}}=0.03$. \textbf{Right:} The performance of a network of $N=18$ cavities with assumed background probability per cavity and per measurement window of $p_{\mathrm{bkg}}=0.15$. 
  }
  \label{fig:NConcidences}
\end{figure}


As a first benchmark, we consider a network of $N = 9$ cavities operated in parallel and assume a background probability per cavity and per measurement window of $p_{\mathrm{bkg}} = 0.03$.
Using the coincidence formalism developed in Sec.~\ref{sec:networkOperation}, we evaluate the expected number of accidental coincidences as a function of the required coincidence order $k$. As shown in Figure~\ref{fig:NConcidences} left, requiring a coincidence of at least $k = 8$ out of $9$ detectors suppresses the expected number of background events to below 0.1 per year, while simultaneously retaining a high signal efficiency. In particular, the probability that a genuine HFGW signal produces a coincidence in at least $8$ detectors exceeds 0.9 for the assumed single-cavity detection efficiency $p_{\rm sig} = 0.9$.

A more challenging scenario is illustrated in Figure~\ref{fig:NConcidences}, right panel, where we assume a significantly larger background probability per site and per measurement window, $p_{\mathrm{bkg}} = 0.15$. In this case, stronger combinatorial suppression is required to reduce the expected number of accidental coincidences below unity over the full data-taking period. As shown in Figure~\ref{fig:NConcidences}, this can be achieved by increasing both the network size and the coincidence requirement. Specifically, for a network of $N = 18$ cavities, requiring coincidences in at least $k = 16$ detectors suffices to suppress the background to an acceptable level while maintaining a high probability of detecting a true signal.

These examples illustrate a central feature of the \GravNet\ concept: as long as the per-site background probability $p_{\mathrm{bkg}}$ is significantly smaller than the single-site detection efficiency $p$, accidental backgrounds can be suppressed to arbitrarily low levels by increasing the number of detectors and imposing appropriate coincidence requirements. In contrast, the signal efficiency is fundamentally limited by the sensitivity of individual detectors.

Consequently, the primary experimental objective must be to maximize the single-cavity detection efficiency, i.e.\ to optimize the detector sensitivity to the smallest possible gravitational-wave strains. Network operation and coincidence analysis then provide a scalable and robust mechanism for background rejection, even in the presence of relatively large per-detector background rates.

\subsection{Expected sensitivity for an advanced readout system and multiple detectors}  
\label{sec:singePhotonDetection}

In the long-term perspective, the high-frequency detectors of \GravNet are envisioned to be equipped with single-photon detectors (SPDs) which are discussed in Sec. \ref{sec:SPD}. In this configuration, the experiment becomes a counting experiment, in which the dominant background contributions arise from thermal photons in the cavity and from intrinsic dark counts of the SPDs. The background count rate of an individual detector is assumed to follow Poisson statistics, with a mean determined by the sum of all noise contributions and proportional to the duration of the detector readout window.
The contribution from thermal (blackbody) radiation inside the cavity can be estimated from the corresponding black body radiation $P_{\rm bb}$, multiplied with the antenna absorption \cite{Dicke}.  At millikelvin temperatures and for GHz-scale frequencies, this thermal contribution is exponentially suppressed and remains negligible at the envisioned operating temperature of $20$\,mK compared to the zero-point energy of the cavity as shown in Fig.\,\ref{fig:noise}.

\begin{figure}
    \centering
    \includegraphics{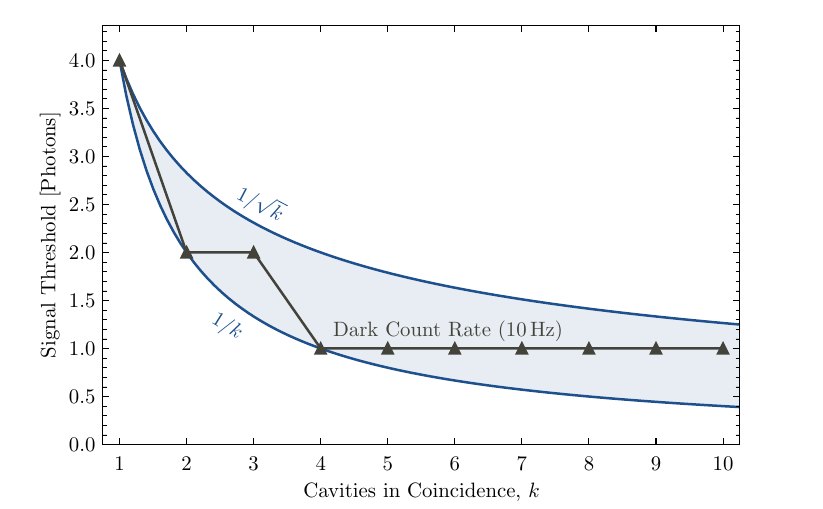}
    \caption{Required minimum threshold of signal photons to achieve a false positive probability per measurement $< 10^{-10}$ as a function of the number of detectors in coincidence. An SPD dark count rate of 10\,Hz is assumed to be the dominant noise source.
    }
    \label{fig:photonCounting_scaling}
\end{figure}

The signal threshold in number of photons $N_{\mathrm{sig}}$ is given by the allowed false positive rate, which is typically chosen to be one per year, and the number of measurement interval within this period. Typical measurement intervals for SPDs are on the order of milliseconds, yielding about $10^{10}$ measurement intervals per year. Correspondingly, the probability of an accidental positive signal detection must be $ < 10^{-10}$ per measurement. 
From this the signal threshold in number of photons $N_{\textrm{thr}}$ is calculated using the inverse survival function of the Poisson distribution with a mean number of background events $\mu_{\textrm{bkg}} = P_{\rm{noise}}$ and the target probability $10^{-10}$. As an example, assuming a dark count rate of 10\,Hz which is close to already achieved values~\cite{Pallegoix:2025dce}, and consequently $P_{\textrm{noise}} = 0.01$, a signal threshold of $N_{\textrm{thr}} = 4$ photons is found. 
Operating multiple detectors in coincidence will drastically reduce the signal threshold, as shown in Sec. \ref{sec:networkOperation}. With three detectors $N_{\textrm{thr}}$ is reduced to the desired target of one photon. Without any further assumptions.
This scaling follows a $1/n$ dependence as shown in Figure\,\ref{fig:photonCounting_scaling}.

With increasing number of detectors in coincidence the detection efficiency of signal photons scales with the binomial sum as shown in eq. \ref{eq:binomSumSignal} in Sec. \ref{sec:networkOperation}. 
To achieve a high overall detection efficiency, one may require a coincidence in $k$ out of $n$ detectors to see a signal. Using four out of seven detectors in coincidence retains the signal threshold of one photon and reaches a detection efficiency of $> 0.95$ assuming the efficiency of a single SPD to be $\epsilon_{\textrm{det}} = 0.8$. Using a four out of 10 coincidence reached a detection efficiency greater than $0.99$. Table~\ref{tab:CoindicendeSetups} summarizes coincidence configurations usable for different assumptions on the dark count rate. This shows that nine detectors with SPDs reach a signal threshold of one photon with dark count rates up to 50\,Hz.

\begin{table}[ht]
    \renewcommand{\arraystretch}{1.0}
    \setlength{\tabcolsep}{4pt}
    \centering
    \begin{tabular*}{0.5\textwidth}{@{\extracolsep{\fill}}cccc}
        \toprule\midrule
        $f_\mathrm{DC}~(p_\mathrm{bkg})$
        & $p_{\mathrm{sig}}^{(k)}>0.90$ & $p_{\mathrm{sig}}^{(k)}>0.95$ & $p_{\mathrm{sig}}^{(k)}>0.99$ \\
        \midrule

        
        
        $\SI{5}{\hertz} ~(0.015)$  & 7/9   & 7/10  & 7/11  \\
        $\SI{10}{\hertz}~(0.03)$   & 8/10  & 8/11  & 9/13  \\
        $\SI{20}{\hertz}~(0.06)$   & 11/14 & 11/14 & 12/17 \\
        
        \midrule\bottomrule
        \end{tabular*}
        
    \caption{
    Minimal coincidence configuration ($k$ out of $n$ cavities) required to reach a false positive rate below $10^{-10}$, corresponding to approximately one background event per year. The rows correspond to different assumptions on the dark count rate $f_\mathrm{DC}$ ($p_\mathrm{bkg}$ --- probability of background photon in a single time window), while the columns indicate the required overall signal detection probability $p_\mathrm{sig}^{(k)}$ of $k$ coincident cavities. A single-cavity signal detection efficiency of $p_\mathrm{sig}=0.9$ is assumed. The coincidence time window used in the calculation is $\SI{3}{\milli\second}$.}
    \label{tab:CoindicendeSetups}
\end{table}

\begin{figure}[t]
    \centering
    \includegraphics{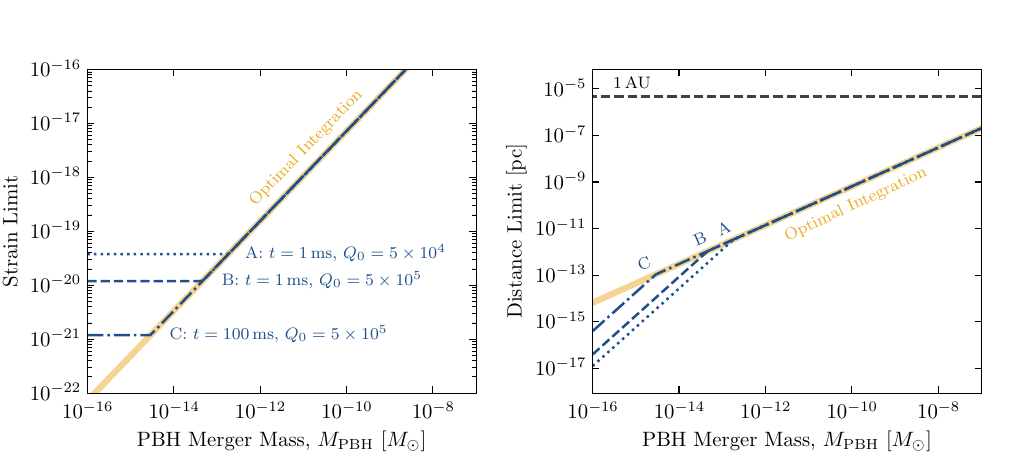}
    \caption{The $95\%$ confidence level limits on the sensitivity assuming a coincidence measurement with single photon detectors, where the number of coincidences is chosen to reach a signal threshold of one detected photon. For the appropriate coincidence configurations refer to Tab.~\ref{tab:CoindicendeSetups}. The coincidence window is taken to be long enough to allow the full transient signal to be included, as shown by the `Optimal Integration' line, computed for $Q_0 = 5\times 10^5$. Shorter coincidence (or readout) windows reduces the sensitivity; in this case, the quality factor becomes important as it changes the time the signal spends within the time window. We take three benchmark configurations to illustrate this (A, B, and C), as indicated by the broken lines.  \textbf{Left:} The detectable strains with PBH merger mass. Lower PBH masses yield longer signals within the bandwidth of the cavity, such that the sensitivity saturates when the signal duration is longer than the coincidence time.  \textbf{Right:} The distance reach after converting the strain limit to a distance for a given PBH merger mass. The broken lines are drawn for the same benchmark points as on the left panel. Also marked is $1\,\mathrm{AU} \simeq 4.8\times10^{-6}\,\mathrm{pc}$.}
    \label{fig:limit-SPD-1photonThreshold}
\end{figure}

 We now turn to the estimate of the sensitivity to PBH mergers.  We require that within the bandwidth of the cavity and within the assumed readout time of 1\,ms (10\,ms) the energy of at least one photon is deposited by the GW in the cavity. Using Eqs.\,\ref{eqn:energy} the required strain is calculated as a function of the PBH merger mass, see Figure\,\ref{fig:limit-SPD-1photonThreshold} (left). 
In this configuration, the strain limit decreases with increasing mass of the signal source. 
For signal durations exceeding the SPD measurement interval, the strain limit becomes independent of the source mass. 
The measurement duration of the SPD and the bandwidth of the cavity determine at which PBH mass and minimum strain the strain limit becomes constant. Increasing the signal threshold allows one to reach lower PBH masses at lower minimum strains, but decreases the sensitivity to fast signals at larger mass values. 

The strain limit is converted to the distance reach, within which a PBH merger would be detected, as shown in Figure\,\ref{fig:limit-SPD-1photonThreshold} (right). 

The limiting factor is the requirement to deposit the energy of at least one photon in the detector. This yields sensitivity independent of the quality factor of the cavity and independent of the readout time of the SPD as long as the time the transient signal spent within the bandwidth of the cavity is smaller than the readout time. If the transients are slow, increased readout time windows increase the sensitivity. Alternatively, adjacent readout windows can be combined in software. 
Comparing these results to the expectations on the sensitivity of the power readout (Figure\,\ref{fig:LinearAmp_DistanceReach}) shows a superior sensitivity for low PBH masses $< 10^{-13} M_\odot$ and sensitivity for large masses $M_{\mathrm{PBH}} > 10^{-10} M_\odot$, which are not accessible with the power readout. 



\clearpage
\section{Conclusion and Outlook}

The \GravNet collaboration aims to establish a novel experimental platform dedicated to the research and development for the detection of high-frequency gravitational waves (HFGWs). The proposed strategy relies on the synchronous operation of multiple, geographically separated electromagnetic cavities immersed in strong magnetic fields, enabling the exploration of detection techniques based on the inverse Gertsenshtein effect in resonant systems. Beyond the search for potential signals, a central objective of \GravNet is the systematic development, validation, and optimization of technologies capable of probing the largely unexplored HFGW domain.

The network configuration plays a crucial role in addressing one of the central experimental challenges: the discrimination of genuine gravitational signals from experimental noise. By exploiting time-coincidence and cross-correlation among distant detectors, the \GravNet network enables robust background rejection, improves the achievable signal-to-noise ratio, and provides a framework for validating the gravitational origin of transient events. In addition, the spatial separation of the detectors allows the reconstruction of arrival times across the network, opening the possibility of constraining the direction of potential sources.

An important feature of the project is the development of resonant cavities capable of tuning over a broad range of frequencies. This capability is particularly relevant for the study of rapidly evolving HFGW signals. While we have explored the promise of such detectors to detect primordial black hole mergers, the theoretical framework developed here indicates that we will have sensitivities to a variety of well-motivated sources.

The \GravNet network is designed to be an open and evolving experimental framework. Additional detector nodes may join the network as the project progresses, further increasing its sensitivity and robustness. Since gravitational waves interact universally with matter, the network can in principle incorporate a variety of detector technologies beyond resonant cavities. This flexibility allows \GravNet to serve as a testbed for exploring different experimental strategies and identifying the most promising approaches to achieve competitive sensitivity in the high-frequency regime.

The \GravNet project will proceed through a phased implementation:
\begin{itemize}
    \item Phase I (Demonstration): Building upon the initial single non-superconducting cavity experiment, this phase will focus on commissioning the first three detectors located in Bonn, Mainz, and Frascati. The primary goal is to validate the time-synchronization and correlation-based data-analysis techniques required for efficient noise rejection and directional sensitivity.
    
    \item Phase II (Network Expansion): Expanding the \GravNet infrastructure to four or more detectors distributed globally. A larger network will enhance coincidence-based background suppression, improve source localization capabilities, and establish a continuous monitoring program for transient HFGW events.
    
    \item Phase III (Sensitivity Goal): Optimizing the network through the deployment of advanced (possibly superconducting) cavities with increased volume ($V$) and magnetic field strength ($B_0$), with the aim of reaching sensitivities capable of probing the primordial black hole dark-matter paradigm in the asteroid mass window and constraining models of primordial gravitational-wave backgrounds.
\end{itemize}

Looking ahead, advances in quantum sensing may provide significant opportunities to further enhance detector performance. In particular, quantum measurement schemes based on entanglement \cite{Brady:2022bus} and distributed quantum sensing \cite{Cai:2025fpe, Wang:2026rwy} could improve the effective sensitivity of spatially separated detectors by enabling correlated measurements across the network. Initiatives such as the Fermilab quantum network IEQNET (Illinois Express Quantum Network) demonstrate the feasibility of long-distance entangled measurements and may offer a technological pathway toward implementing quantum-enhanced gravitational-wave detection strategies in the future.

By systematically exploring the largely uncharted HFGW spectrum and fostering the development of new experimental techniques, \GravNet represents an important step toward opening a new observational window into the Universe. Whether through direct detection of high-frequency gravitational waves or through the technological advances developed along the way, the program has the potential to significantly deepen our understanding of gravity, the nature of dark matter, and the physics of the early Universe.

\section*{Acknowledgments}
This research is primarily supported by the ERC-2024-SYG 101167211 grant (\GravNet DOI: 10.3030/101167211), as well as by the
 Cluster of Excellence ``Precision Physics, Fundamental Interactions, and Structure of Matter'' (PRISMA++ EXC 2118/2) funded by the German Research Foundation (DFG) within the German Excellence Strategy (Project ID 390831469), and by the COST Action within the project COSMIC WISPers (Grant No. CA21106). Views and opinions expressed are however those of the author(s) only and do not necessarily reflect those of the European Union or the European Research Council Executive Agency. Neither the European Union nor the granting authority can be held responsible for them. 

 This publication is part of the grant PID2023-146686NB-C31 funded by MICIU/AEI/10.13039/501100011033/ and by FEDER, UE. IFAE is partially funded by the CERCA program of the Generalitat de Catalunya.   D.B. acknowledges the support from the European Research Area (ERA) via the UNDARK project of the Widening participation and spreading excellence programme (project number 101159929). L.V.\ acknowledges support by Istituto Nazionale di Fisica Nucleare (INFN) through the Commissione Scientifica Nazionale 4 (CSN4) Iniziativa Specifica ``Quantum Universe'' (QGSKY). J.G.\ is funded by Grant No.\ CNS2023-143767, funded by MICIU/AEI/10.13039/501100011033 and by European Union NextGenerationEU/PRTR. 

\bibliographystyle{apsrev4-1} 
\bibliography{./Bibliography}

@article{Gertsenshtein:1962kfm,
    author = "Gertsenshtein, M. E. and Pustovoit, V. I.",
    title = "{On the Detection of Low Frequency Gravitational Waves}",
    journal = "Sov. Phys. JETP",
    volume = "16",
    pages = "433",
    year = "1962"
}

@article{Carr:2020gox,
    author = "Carr, Bernard and Kohri, Kazunori and Sendouda, Yuuiti and Yokoyama, Jun'ichi",
    title = "{Constraints on primordial black holes}",
    eprint = "2002.12778",
    archivePrefix = "arXiv",
    primaryClass = "astro-ph.CO",
    reportNumber = "RESCEU-03/20; KEK-Cosmo-249; KEK-TH-2199; IPMU20-0024",
    doi = "10.1088/1361-6633/ac1e31",
    journal = "Rept. Prog. Phys.",
    volume = "84",
    number = "11",
    pages = "116902",
    year = "2021"
}

@article{Witte:2024drg,
    author = "Witte, Samuel J. and Mummery, Andrew",
    title = "{Stepping up superradiance constraints on axions}",
    eprint = "2412.03655",
    archivePrefix = "arXiv",
    primaryClass = "hep-ph",
    doi = "10.1103/PhysRevD.111.083044",
    journal = "Phys. Rev. D",
    volume = "111",
    number = "8",
    pages = "083044",
    year = "2025"
}

@article{Giudice:2016zpa,
    author = "Giudice, Gian F. and McCullough, Matthew and Urbano, Alfredo",
    title = "{Hunting for Dark Particles with Gravitational Waves}",
    eprint = "1605.01209",
    archivePrefix = "arXiv",
    primaryClass = "hep-ph",
    reportNumber = "CERN-TH-2016-106",
    doi = "10.1088/1475-7516/2016/10/001",
    journal = "JCAP",
    volume = "10",
    pages = "001",
    year = "2016"
}

@article{Green:2020jor,
    author = "Green, Anne M. and Kavanagh, Bradley J.",
    title = "{Primordial Black Holes as a dark matter candidate}",
    eprint = "2007.10722",
    archivePrefix = "arXiv",
    primaryClass = "astro-ph.CO",
    doi = "10.1088/1361-6471/abc534",
    journal = "J. Phys. G",
    volume = "48",
    number = "4",
    pages = "043001",
    year = "2021"
}

@article{Goryachev:2014yra,
    author = "Goryachev, Maxim and Tobar, Michael E.",
    title = "{Gravitational Wave Detection with High Frequency Phonon Trapping Acoustic Cavities}",
    eprint = "1410.2334",
    archivePrefix = "arXiv",
    primaryClass = "gr-qc",
    reportNumber = "Erratum: Phys. Rev. D, 108, 129901(E) (2023)",
    doi = "10.1103/PhysRevD.90.102005",
    journal = "Phys. Rev. D",
    volume = "90",
    number = "10",
    pages = "102005",
    year = "2014",
    note = "[Erratum: Phys.Rev.D 108, 129901 (2023)]"
}

@article{Jungkind:2025oqm,
    author = "Jungkind, Christopher M. and Seymour, Brian C. and Laeuger, Andrew and Chen, Yanbei",
    title = "{Prospects for high-frequency gravitational-wave detection with GEO600}",
    eprint = "2506.08315",
    archivePrefix = "arXiv",
    primaryClass = "gr-qc",
    doi = "10.1103/y3h1-lqy3",
    journal = "Phys. Rev. D",
    volume = "113",
    number = "2",
    pages = "024057",
    year = "2026"
}

@article{Schnabel:2024hem,
    author = "Schnabel, Roman and Korobko, Mikhail",
    title = "{Optical sensitivities of current gravitational wave observatories at higher kHz, MHz and GHz frequencies}",
    eprint = "2409.03019",
    archivePrefix = "arXiv",
    primaryClass = "astro-ph.IM",
    doi = "10.1038/s41598-025-08668-x",
    journal = "Sci. Rep.",
    volume = "15",
    number = "1",
    pages = "25733",
    year = "2025"
}

@article{Chaudhuri:2019ntz,
    author = "Chaudhuri, Saptarshi and Irwin, Kent D. and Graham, Peter W. and Mardon, Jeremy",
    title = "{Optimal Electromagnetic Searches for Axion and Hidden-Photon Dark Matter}",
    eprint = "1904.05806",
    archivePrefix = "arXiv",
    primaryClass = "hep-ex",
    month = "4",
    year = "2019"
}

@article{Arvanitaki:2009fg,
    author = "Arvanitaki, Asimina and Dimopoulos, Savas and Dubovsky, Sergei and Kaloper, Nemanja and March-Russell, John",
    title = "{String Axiverse}",
    eprint = "0905.4720",
    archivePrefix = "arXiv",
    primaryClass = "hep-th",
    doi = "10.1103/PhysRevD.81.123530",
    journal = "Phys. Rev. D",
    volume = "81",
    pages = "123530",
    year = "2010"
}

@article{Ringwald:2022xif,
    author = "Ringwald, Andreas and Tamarit, Carlos",
    title = "{Revealing the cosmic history with gravitational waves}",
    eprint = "2203.00621",
    archivePrefix = "arXiv",
    primaryClass = "hep-ph",
    reportNumber = "DESY 22-037, TUM-HEP-1389-22",
    doi = "10.1103/PhysRevD.106.063027",
    journal = "Phys. Rev. D",
    volume = "106",
    number = "6",
    pages = "063027",
    year = "2022"
}

@article{Brito:2015oca,
    author = "Brito, Richard and Cardoso, Vitor and Pani, Paolo",
    title = "{Superradiance}: {New Frontiers in Black Hole
Physics}",
    eprint = "1501.06570",
    archivePrefix = "arXiv",
    primaryClass = "gr-qc",
    doi = "10.1007/978-3-319-19000-6",
    journal = "Lect. Notes Phys.",
    volume = "906",
    pages = "pp.1--237",
    year = "2015"
}

@article{Renzini:2022alw,
    author = "Renzini, Arianna I. and Goncharov, Boris and Jenkins, Alexander C. and Meyers, Pat M.",
    title = "{Stochastic Gravitational-Wave Backgrounds: Current Detection Efforts and Future Prospects}",
    eprint = "2202.00178",
    archivePrefix = "arXiv",
    primaryClass = "gr-qc",
    doi = "10.3390/galaxies10010034",
    journal = "Galaxies",
    volume = "10",
    number = "1",
    pages = "34",
    year = "2022"
}

@article{Janssen:2014dka,
    author = "Janssen, Gemma and others",
    editor = "Bourke, Tyler L. and others",
    title = "{Gravitational wave astronomy with the SKA}",
    eprint = "1501.00127",
    archivePrefix = "arXiv",
    primaryClass = "astro-ph.IM",
    doi = "10.22323/1.215.0037",
    journal = "PoS",
    volume = "AASKA14",
    pages = "037",
    year = "2015"
}

@article{Ferreira:2020fam,
    author = "Ferreira, Elisa G. M.",
    title = "{Ultra-light dark matter}",
    eprint = "2005.03254",
    archivePrefix = "arXiv",
    primaryClass = "astro-ph.CO",
    doi = "10.1007/s00159-021-00135-6",
    journal = "Astron. Astrophys. Rev.",
    volume = "29",
    number = "1",
    pages = "7",
    year = "2021"
}

@article{Maggiore:2019uih,
    author = "Maggiore, Michele and others",
    title = "{Science Case for the Einstein Telescope}",
    eprint = "1912.02622",
    archivePrefix = "arXiv",
    primaryClass = "astro-ph.CO",
    doi = "10.1088/1475-7516/2020/03/050",
    journal = "JCAP",
    volume = "03",
    pages = "050",
    year = "2020"
}

@ARTICLE{LISA,
       author = {{Amaro-Seoane}, Pau and others},
        title = "{Laser Interferometer Space Antenna}",
      journal = {arXiv e-prints},
         year = 2017,
        month = feb,
          eid = {arXiv:1702.00786},
        pages = {arXiv:1702.00786},
          doi = {10.48550/arXiv.1702.00786},
archivePrefix = {arXiv},
       eprint = {1702.00786},
 primaryClass = {astro-ph.IM},
       adsurl = {https://ui.adsabs.harvard.edu/abs/2017arXiv170200786A}
}

@article{Pallegoix:2025dce,
    author = "Pallegoix, Louis and Travesedo, Jaime and May, Alexandre S. and Balembois, L{\'e}o and Vion, Denis and Bertet, Patrice and Flurin, Emmanuel",
    title = "{Enhancing the sensitivity of single microwave photon detection with bandwidth tunability}",
    eprint = "2501.07354",
    archivePrefix = "arXiv",
    primaryClass = "quant-ph",
    month = "1",
    year = "2025"
}

@article{Vagnozzi:2022qmc,
    author = "Vagnozzi, Sunny and Loeb, Abraham",
    title = "{The Challenge of Ruling Out Inflation via the Primordial Graviton Background}",
    eprint = "2208.14088",
    archivePrefix = "arXiv",
    primaryClass = "astro-ph.CO",
    doi = "10.3847/2041-8213/ac9b0e",
    journal = "Astrophys. J. Lett.",
    volume = "939",
    number = "2",
    pages = "L22",
    year = "2022"
}

@article{Lawson:2019brd,
    author = "Lawson, Matthew and Millar, Alexander J. and Pancaldi, Matteo and Vitagliano, Edoardo and Wilczek, Frank",
    title = "{Tunable axion plasma haloscopes}",
    eprint = "1904.11872",
    archivePrefix = "arXiv",
    primaryClass = "hep-ph",
    reportNumber = "NORDITA-2019-038, MIT-CTP-5116, MPP-2019-83",
    doi = "10.1103/PhysRevLett.123.141802",
    journal = "Phys. Rev. Lett.",
    volume = "123",
    number = "14",
    pages = "141802",
    year = "2019"
}

@article{Franciolini:2022htd,
    author = "Franciolini, Gabriele and Maharana, Anshuman and Muia, Francesco",
    title = "{Hunt for light primordial black hole dark matter with ultrahigh-frequency gravitational waves}",
    eprint = "2205.02153",
    archivePrefix = "arXiv",
    primaryClass = "astro-ph.CO",
    doi = "10.1103/PhysRevD.106.103520",
    journal = "Phys. Rev. D",
    volume = "106",
    number = "10",
    pages = "103520",
    year = "2022"
}

@article{Carr:2021bzv,
    author = "Carr, Bernard and Kuhnel, Florian",
    title = "{Primordial black holes as dark matter candidates}",
    eprint = "2110.02821",
    archivePrefix = "arXiv",
    primaryClass = "astro-ph.CO",
    doi = "10.21468/SciPostPhysLectNotes.48",
    journal = "SciPost Phys. Lect. Notes",
    volume = "48",
    pages = "1",
    year = "2022"
}

@book{Maggiore:2007ulw,
    author = "Maggiore, Michele",
    title = "{Gravitational Waves. Vol. 1: Theory and Experiments}",
    doi = "10.1093/acprof:oso/9780198570745.001.0001",
    isbn = "978-0-19-171766-6, 978-0-19-852074-0",
    publisher = "Oxford University Press",
    year = "2007"
}

@article{KAGRA:2013rdx,
    author = "Abbott, B. P. and others",
    collaboration = "KAGRA, LIGO Scientific, Virgo, VIRGO",
    title = "{Prospects for observing and localizing gravitational-wave transients with Advanced LIGO, Advanced Virgo and KAGRA}",
    eprint = "1304.0670",
    archivePrefix = "arXiv",
    primaryClass = "gr-qc",
    reportNumber = "LIGO-P1200087, VIR-0288A-12, JGW-P1808427",
    doi = "10.1007/s41114-020-00026-9",
    journal = "Living Rev. Rel.",
    volume = "21",
    number = "1",
    pages = "3",
    year = "2018"
}

@misc{CAPPCavity,
    title = {CAPP: Superconducting Cavity for Dark Matter Axion Search},
    howpublished = {\url{https://agenda.infn.it/event/20431/contributions/137693/attachments/82473/108372/2021_June_PATRAS_Meeting_Danho_Ahn_v5.pdf}},
    year = "2021"
}

@article{Berlin:2023grv,
    author = {Berlin, Asher and Blas, Diego and Tito D'Agnolo, Raffaele and Ellis, Sebastian A. R. and Harnik, Roni and Kahn, Yonatan and Sch\"utte-Engel, Jan and Wentzel, Michael},
    title = "{Electromagnetic cavities as mechanical bars for gravitational waves}",
    eprint = "2303.01518",
    archivePrefix = "arXiv",
    primaryClass = "hep-ph",
    reportNumber = "FERMILAB-PUB-22-892-SQMS-T",
    doi = "10.1103/PhysRevD.108.084058",
    journal = "Phys. Rev. D",
    volume = "108",
    number = "8",
    pages = "084058",
    year = "2023"
}

@article{Clarke:2020bil,
    author = "Clarke, Thomas J. and Copeland, Edmund J. and Moss, Adam",
    title = "{Constraints on primordial gravitational waves from the Cosmic Microwave Background}",
    eprint = "2004.11396",
    archivePrefix = "arXiv",
    primaryClass = "astro-ph.CO",
    doi = "10.1088/1475-7516/2020/10/002",
    journal = "JCAP",
    volume = "10",
    pages = "002",
    year = "2020",
}

@article{LIGOScientific:2016aoc,
    author = "Abbott, B. P. and others",
    collaboration = "LIGO Scientific, Virgo",
    title = "{Observation of Gravitational Waves from a Binary Black Hole Merger}",
    eprint = "1602.03837",
    archivePrefix = "arXiv",
    primaryClass = "gr-qc",
    reportNumber = "LIGO-P150914",
    doi = "10.1103/PhysRevLett.116.061102",
    journal = "Phys. Rev. Lett.",
    volume = "116",
    number = "6",
    pages = "061102",
    year = "2016"
}

@article{Aybas:2021nvn,
    author = "Aybas, Deniz and others",
    title = "{Search for Axionlike Dark Matter Using Solid-State Nuclear Magnetic Resonance}",
    eprint = "2101.01241",
    archivePrefix = "arXiv",
    primaryClass = "hep-ex",
    doi = "10.1103/PhysRevLett.126.141802",
    journal = "Phys. Rev. Lett.",
    volume = "126",
    number = "14",
    pages = "141802",
    year = "2021"
}

@article{Alesini:2023qed,
    author = "Alesini, David and others",
    title = "{The future search for low-frequency axions and new physics with the FLASH resonant cavity experiment at Frascati National Laboratories}",
    eprint = "2309.00351",
    archivePrefix = "arXiv",
    primaryClass = "physics.ins-det",
    reportNumber = "CA21106; CA21136",
    doi = "10.1016/j.dark.2023.101370",
    journal = "Phys. Dark Univ.",
    volume = "42",
    pages = "101370",
    year = "2023"
}

@article{NANOGrav:2023gor,
    author = "Agazie, Gabriella and others",
    collaboration = "NANOGrav",
    title = "{The NANOGrav 15 yr Data Set: Evidence for a Gravitational-wave Background}",
    eprint = "2306.16213",
    archivePrefix = "arXiv",
    primaryClass = "astro-ph.HE",
    doi = "10.3847/2041-8213/acdac6",
    journal = "Astrophys. J. Lett.",
    volume = "951",
    number = "1",
    pages = "L8",
    year = "2023"
}

@article{Schneemann:2023bqc,
    author = "Schneemann, Tim and Schmieden, Kristof and Schott, Matthias",
    title = "{First results of the SUPAX Experiment: Probing Dark Photons}",
    eprint = "2308.08337",
    archivePrefix = "arXiv",
    primaryClass = "hep-ex",
    month = "8",
    year = "2023"
}

@article{ADMX:2010ubl,
    author = "Wagner, A. and others",
    collaboration = "ADMX",
    title = "{A Search for Hidden Sector Photons with ADMX}",
    eprint = "1007.3766",
    archivePrefix = "arXiv",
    primaryClass = "hep-ex",
    doi = "10.1103/PhysRevLett.105.171801",
    journal = "Phys. Rev. Lett.",
    volume = "105",
    pages = "171801",
    year = "2010"
}

@article{Roberts:2017hla,
    author = "Roberts, Benjamin M. and Blewitt, Geoffrey and Dailey, Conner and Murphy, Mac and Pospelov, Maxim and Rollings, Alex and Sherman, Jeff and Williams, Wyatt and Derevianko, Andrei",
    title = "{Search for domain wall dark matter with atomic clocks on board global positioning system satellites}",
    eprint = "1704.06844",
    archivePrefix = "arXiv",
    primaryClass = "hep-ph",
    doi = "10.1038/s41467-017-01440-4",
    journal = "Nature Commun.",
    volume = "8",
    number = "1",
    pages = "1195",
    year = "2017"
}

@article{Berlin:2021txa,
    author = {Berlin, Asher and Blas, Diego and Tito D'Agnolo, Raffaele and Ellis, Sebastian A. R. and Harnik, Roni and Kahn, Yonatan and Sch\"utte-Engel, Jan},
    title = "{Detecting high-frequency gravitational waves with microwave cavities}",
    eprint = "2112.11465",
    archivePrefix = "arXiv",
    primaryClass = "hep-ph",
    reportNumber = "FERMILAB-PUB-21-724-SQMS-T",
    doi = "10.1103/PhysRevD.105.116011",
    journal = "Phys. Rev. D",
    volume = "105",
    number = "11",
    pages = "116011",
    year = "2022"
}

@article{Ballantini:2005am,
    author = "Ballantini, R. and others",
    title = "{Microwave apparatus for gravitational waves observation}",
    eprint = "gr-qc/0502054",
    archivePrefix = "arXiv",
    reportNumber = "INFN-TC-05-05",
    month = "2",
    year = "2005"
}

@article{Aggarwal:2020olq,
    author = "Aggarwal, Nancy and others",
    title = "{Challenges and opportunities of gravitational-wave searches at MHz to GHz frequencies}",
    eprint = "2011.12414",
    archivePrefix = "arXiv",
    primaryClass = "gr-qc",
    reportNumber = "CERN-TH-2020-185, HIP-2020-28/TH, DESY 20-195, CERN-TH-2020-185, HIP-2020-28/TH, DESY 20-195",
    doi = "10.1007/s41114-021-00032-5",
    journal = "Living Rev. Rel.",
    volume = "24",
    number = "1",
    pages = "4",
    year = "2021"
}

@article{Heisig:2025oim,
    author = "Heisig, Jan",
    title = "{Resonant Loop Interferometers for High-Frequency Gravitational Waves}",
    eprint = "2510.13957",
    archivePrefix = "arXiv",
    primaryClass = "gr-qc",
    month = "10",
    year = "2025"
}

@article{Blas:2022xco,
    author = "Blas, Diego and Casalderrey-Solana, Jorge and Mateos, David and Sanchez-Garitaonandia, Mikel",
    title = "{Megahertz Gravitational Waves from Neutron Star Mergers}",
    eprint = "2210.03171",
    archivePrefix = "arXiv",
    primaryClass = "hep-th",
    doi = "10.1103/6yz9-94ql",
    journal = "Phys. Rev. Lett.",
    volume = "136",
    number = "10",
    pages = "101401",
    year = "2026"
}

@article{Alesini:2022lnp,
    author = "Alesini, D. and others",
    title = "{Search for Galactic axions with a high-Q dielectric cavity}",
    eprint = "2208.12670",
    archivePrefix = "arXiv",
    primaryClass = "hep-ex",
    doi = "10.1103/PhysRevD.106.052007",
    journal = "Phys. Rev. D",
    volume = "106",
    number = "5",
    pages = "052007",
    year = "2022"
}

@article{Gramolin2021search,
  title={Search for axion-like dark matter with ferromagnets},
  author={Gramolin, Alexander V and Aybas, Deniz and Johnson, Dorian and Adam, Janos and Sushkov, Alexander O},
  journal={Nature Physics},
  volume={17},
  number={1},
  pages={79--84},
  year={2021},
  publisher={Nature Publishing Group UK London}
}

@article{Tretiak2022_RFDMPRL,
	author = {Tretiak, Oleg and Zhang, Xue and Figueroa, Nataniel L. and Antypas, Dionysios and Brogna, Andrea and Banerjee, Abhishek and Perez, Gilad and Budker, Dmitry},
	doi = {10.1103/PhysRevLett.129.031301},
	issue = {3},
	journal = {Phys. Rev. Lett.},
	month = {Jul},
	numpages = {7},
	pages = {031301},
	publisher = {American Physical Society},
	title = {Improved Bounds on Ultralight Scalar Dark Matter in the Radio-Frequency Range},
	url = {https://link.aps.org/doi/10.1103/PhysRevLett.129.031301},
	volume = {129},
	year = {2022}}

@article{Afach2021_GNOME_NP,
	author = {Afach, Samer and Buchler, Ben C. and Budker, Dmitry and Dailey, Conner and Derevianko, Andrei and Dumont, Vincent and Figueroa, Nataniel L. and Gerhardt, Ilja and Gruji{\'c}, Zoran D. and Guo, Hong and Hao, Chuanpeng and Hamilton, Paul S. and Hedges, Morgan and Jackson Kimball, Derek F. and Kim, Dongok and Khamis, Sami and Kornack, Thomas and Lebedev, Victor and Lu, Zheng-Tian and Masia-Roig, Hector and Monroy, Madeline and Padniuk, Mikhail and Palm, Christopher A. and Park, Sun Yool and Paul, Karun V. and Penaflor, Alexander and Peng, Xiang and Pospelov, Maxim and Preston, Rayshaun and Pustelny, Szymon and Scholtes, Theo and Segura, Perrin C. and Semertzidis, Yannis K. and Sheng, Dong and Shin, Yun Chang and Smiga, Joseph A. and Stalnaker, Jason E. and Sulai, Ibrahim and Tandon, Dhruv and Wang, Tao and Weis, Antoine and Wickenbrock, Arne and Wilson, Tatum and Wu, Teng and Wurm, David and Xiao, Wei and Yang, Yucheng and Yu, Dongrui and Zhang, Jianwei},
	da = {2021/12/07},
	doi = {10.1038/s41567-021-01393-y},
	id = {Afach2021},
	isbn = {1745-2481},
	journal = {Nature Physics},
	title = {Search for topological defect dark matter with a global network of optical magnetometers},
	ty = {JOUR},
    volume = {17},
    pages = {1396},
	url = {https://doi.org/10.1038/s41567-021-01393-y},
	year = {2021}}

@article{Afach2023WhatCanGNOME,
	author = {Afach, Samer and Aybas Tumturk, Deniz and Bekker, Hendrik and Buchler, Ben C. and Budker, Dmitry and Cervantes, Kaleb and Derevianko, Andrei and Eby, Joshua and Figueroa, Nataniel L. and Folman, Ron and Gavil{\'a}n-Mart{\'\i}n, Daniel and Givon, Menachem and Gruji{\'c}, Zoran D. and Guo, Hong and Hamilton, Paul and Hedges, Morgan P. and Jackson Kimball, Derek F. and Khamis, Sami and Kim, Dongok and Klinger, Emmanuel and Kryemadhi, Abaz and Liu, Xiyu and {\L}ukasiewicz, Grzegorz and Masia-Roig, Hector and Padniuk, Mikhail and Palm, Christopher A. and Park, Sun Yool and Pearson, Heather R. and Peng, Xiang and Pospelov, Maxim and Pustelny, Szymon and Rosenzweig, Yossi and Ruimi, Ophir M. and Scholtes, Theo and Segura, Perrin C. and Semertzidis, Yannis K. and Shin, Yun Chang and Smiga, Joseph A. and Stadnik, Yevgeny V. and Stalnaker, Jason E. and Sulai, Ibrahim A. and Tandon, Dhruv and Vu, Kenneth and Weis, Antoine and Wickenbrock, Arne and Wilson, Tatum Z. and Wu, Teng and Xiao, Wei and Yang, Yucheng and Yu, Dongrui and Yu, Felix and Zhang, Jianwei and Zhao, Yixin},
	doi = {https://doi.org/10.1002/andp.202300083},
	eprint = {https://onlinelibrary.wiley.com/doi/pdf/10.1002/andp.202300083},
	journal = {Annalen der Physik},
    volume = {536},
    number = {1},
	pages = {2300083},
	title = {What Can a GNOME Do? Search Targets for the Global Network of Optical Magnetometers for Exotic Physics Searches},
    year = {2024},
	url = {https://onlinelibrary.wiley.com/doi/abs/10.1002/andp.202300083}}

@article{Berlin:2022hfx,
    author = "Berlin, Asher and others",
    title = "{Searches for New Particles, Dark Matter, and Gravitational Waves with SRF Cavities}",
    eprint = "2203.12714",
    archivePrefix = "arXiv",
    primaryClass = "hep-ph",
    reportNumber = "FERMILAB-PUB-22-150-SQMS-T",
    month = "3",
    year = "2022"
}

@article{DEliaCBJJ,
    author = "D'Elia, Alessandro and others",
    title = "{Stepping Closer to Pulsed Single Microwave Photon Detectors for Axions Search}",
    eprint = "2302.07556",
    archivePrefix = "arXiv",
    primaryClass = "quant-ph",
    doi = "10.1109/TASC.2022.3218072",
    journal = "IEEE Trans. Appl. Supercond.",
    volume = "33",
    number = "1",
    pages = "1500109",
    year = "2023"
}

@article{AlimentiSCCavity,
    author = "Alimenti, Andrea and Torokhtii, Kostiantyn and Di Gioacchino, Daniele and Gatti, Claudio and Silva, Enrico and Pompeo, Nicola",
    title = "{Impact of Superconductors\textquoteright{} Properties on the Measurement Sensitivity of Resonant-Based Axion Detectors}",
    eprint = "2112.12775",
    archivePrefix = "arXiv",
    primaryClass = "physics.ins-det",
    doi = "10.3390/instruments6010001",
    journal = "Instruments",
    volume = "6",
    number = "1",
    pages = "1",
    year = "2021"
}

@article{PhysRevApplied.17.L061005,
  title = {Biaxially Textured ${\mathrm{YBa}}_{2}{\mathrm{Cu}}_{3}{\mathrm{O}}_{7\ensuremath{-}x}$ Microwave Cavity in a High Magnetic Field for a Dark-Matter Axion Search},
  author = {Ahn, Danho and others},
  journal = {Phys. Rev. Appl.},
  volume = {17},
  issue = {6},
  pages = {L061005},
  numpages = {6},
  year = {2022},
  month = {Jun},
  publisher = {American Physical Society},
  doi = {10.1103/PhysRevApplied.17.L061005},
  url = {https://link.aps.org/doi/10.1103/PhysRevApplied.17.L061005}
}

@article{QUAXNBTI,
    author = "Alesini, D. and others",
    title = "{Galactic axions search with a superconducting resonant cavity}",
    eprint = "1903.06547",
    archivePrefix = "arXiv",
    primaryClass = "physics.ins-det",
    doi = "10.1103/PhysRevD.99.101101",
    journal = "Phys. Rev. D",
    volume = "99",
    number = "10",
    pages = "101101",
    year = "2019"
}

@article{KuzminCBJJ,
    author = "Kuzmin, Leonid S. and Sobolev, Alexander S. and Gatti, Claudio and Di Gioacchino, Daniele and Crescini, Nicol\`o and Gordeeva, Anna and Il'ichev, Eudeni",
    title = "{Single Photon Counter based on a Josephson Junction at 14 GHz for searching Galactic Axions}",
    doi = "10.1109/TASC.2018.2850019",
    journal = "IEEE Trans. Appl. Supercond.",
    volume = "28",
    number = "7",
    pages = "2400505",
    year = "2018"
}

@ARTICLE{Supergalax,
  author={Gatti, C. and others},
  journal={IEEE Transactions on Applied Superconductivity}, 
  title={Coherent Quantum Network of Superconducting Qubits as a Highly Sensitive Detector of Microwave Photons for Searching of Galactic Axions}, 
  year={2023},
  volume={33},
  number={5},
  pages={1-5},
  doi={10.1109/TASC.2023.3263807}}

@article{Roch,
  title = {Widely Tunable, Nondegenerate Three-Wave Mixing Microwave Device Operating near the Quantum Limit},
  author = {Roch, N. and Flurin, E. and Nguyen, F. and Morfin, P. and Campagne-Ibarcq, P. and Devoret, M. H. and Huard, B.},
  journal = {Phys. Rev. Lett.},
  volume = {108},
  issue = {14},
  pages = {147701},
  numpages = {5},
  year = {2012},
  month = {Apr},
  publisher = {American Physical Society},
  doi = {10.1103/PhysRevLett.108.147701},
  url = {https://link.aps.org/doi/10.1103/PhysRevLett.108.147701}
}

@article{Zhong,
doi = {10.1088/1367-2630/15/12/125013},
url = {https://dx.doi.org/10.1088/1367-2630/15/12/125013},
year = {2013},
month = {dec},
publisher = {IOP Publishing},
volume = {15},
number = {12},
pages = {125013},
author = {L Zhong and E P Menzel and R Di Candia and P Eder and M Ihmig and A Baust and M Haeberlein and E Hoffmann and K Inomata and T Yamamoto and Y Nakamura and E Solano and F Deppe and A Marx and R Gross},
title = {Squeezing with a flux-driven Josephson parametric amplifier},
journal = {New Journal of Physics},
}

@article{Backes,
author ={Backes, K.M. and Palken, D.A. and Kenany, S.A. and others}, title={A quantum enhanced search for dark matter axions},
journal={Nature},
volume={590},
pages={238–242},
year={2021}
}

@article{Brady,
  title = {Entangled Sensor-Networks for Dark-Matter Searches},
  author = {Brady, Anthony J. and Gao, Christina and Harnik, Roni and Liu, Zhen and Zhang, Zheshen and Zhuang, Quntao},
  journal = {PRX Quantum},
  volume = {3},
  issue = {3},
  pages = {030333},
  numpages = {33},
  year = {2022},
  month = {Sep},
  publisher = {American Physical Society},
  doi = {10.1103/PRXQuantum.3.030333},
  url = {https://link.aps.org/doi/10.1103/PRXQuantum.3.030333}
}

@article{Kim:2025izt,
    author = "Kim, Younggeun and others",
    title = "{Search for high-frequency gravitational waves via re-analysis of cavity axion data}",
    eprint = "2511.17817",
    archivePrefix = "arXiv",
    primaryClass = "hep-ex",
    month = "11",
    year = "2025"
}

@article{Borysenkova:2026tbd,
    author = "Borysenkova, Yuliia and Kvietkauskas, Tomas and  Amaral, Dorian and Kalia, Saarik and Blas, Diego",
    title = "{GravNet: Waveforms and Cavities}",
    eprint = "2603.xxxx",
    archivePrefix = "arXiv",
    primaryClass = "hep-ph",
    month = "1",
    year = "2026"
}

@article{Gatti:2024mde,
    author = "Gatti, Claudio and Visinelli, Luca and Zantedeschi, Michael",
    title = "{Cavity detection of gravitational waves: Where do we stand?}",
    eprint = "2403.18610",
    archivePrefix = "arXiv",
    primaryClass = "gr-qc",
    reportNumber = "CA21106; CA21136",
    doi = "10.1103/PhysRevD.110.023018",
    journal = "Phys. Rev. D",
    volume = "110",
    number = "2",
    pages = "023018",
    year = "2024"
}

@article{Gue:2026kga,
    author = "Gu{\'e}, Jordan and Krokotsch, Tom and Moortgat-Pick, Gudrid",
    title = "{Covariant eigenmode overlap formalism for gravitational wave signals in electromagnetic cavities}",
    eprint = "2602.08507",
    archivePrefix = "arXiv",
    primaryClass = "gr-qc",
    month = "2",
    year = "2026"
}

@article{Baryakhtar:2020gao,
    author = "Baryakhtar, Masha and Galanis, Marios and Lasenby, Robert and Simon, Olivier",
    title = "{Black hole superradiance of self-interacting scalar fields}",
    eprint = "2011.11646",
    archivePrefix = "arXiv",
    primaryClass = "hep-ph",
    doi = "10.1103/PhysRevD.103.095019",
    journal = "Phys. Rev. D",
    volume = "103",
    number = "9",
    pages = "095019",
    year = "2021"
}

@article{Ghiglieri:2020mhm,
    author = "Ghiglieri, J. and Jackson, G. and Laine, M. and Zhu, Y.",
    title = "{Gravitational wave background from Standard Model physics: Complete leading order}",
    eprint = "2004.11392",
    archivePrefix = "arXiv",
    primaryClass = "hep-ph",
    doi = "10.1007/JHEP07(2020)092",
    journal = "JHEP",
    volume = "07",
    pages = "092",
    year = "2020"
}

@article{Arvanitaki:2010sy,
    author = "Arvanitaki, Asimina and Dubovsky, Sergei",
    title = "{Exploring the String Axiverse with Precision Black Hole Physics}",
    eprint = "1004.3558",
    archivePrefix = "arXiv",
    primaryClass = "hep-th",
    doi = "10.1103/PhysRevD.83.044026",
    journal = "Phys. Rev. D",
    volume = "83",
    pages = "044026",
    year = "2011"
}

@article{Maccone,
  title = {Quantum Metrology of Noisy Spreading Channels},
  author = {G\'orecki, Wojciech and Riccardi, Alberto and Maccone, Lorenzo},
  journal = {Phys. Rev. Lett.},
  volume = {129},
  issue = {24},
  pages = {240503},
  numpages = {9},
  year = {2022},
  month = {Dec},
  publisher = {American Physical Society},
  doi = {10.1103/PhysRevLett.129.240503},
  url = {https://link.aps.org/doi/10.1103/PhysRevLett.129.240503}
}

@article{Schuster,
author ={D. Schuster, A. Houck, J. Schreier and others}, 
title={Resolving photon number states in a superconducting circuit},
journal={Nature},
volume={445},
pages={515-518},
year={2007}
}

@article{Koch,
  title = {Charge-insensitive qubit design derived from the Cooper pair box},
  author = {Koch, Jens and Yu, Terri M. and Gambetta, Jay and Houck, A. A. and Schuster, D. I. and Majer, J. and Blais, Alexandre and Devoret, M. H. and Girvin, S. M. and Schoelkopf, R. J.},
  journal = {Phys. Rev. A},
  volume = {76},
  issue = {4},
  pages = {042319},
  numpages = {19},
  year = {2007},
  month = {Oct},
  publisher = {American Physical Society},
  doi = {10.1103/PhysRevA.76.042319},
  url = {https://link.aps.org/doi/10.1103/PhysRevA.76.042319}
}

@article{Dixit,
  title = {Searching for Dark Matter with a Superconducting Qubit},
  author = {Dixit, Akash V. and Chakram, Srivatsan and He, Kevin and Agrawal, Ankur and Naik, Ravi K. and Schuster, David I. and Chou, Aaron},
  journal = {Phys. Rev. Lett.},
  volume = {126},
  issue = {14},
  pages = {141302},
  numpages = {7},
  year = {2021},
  month = {Apr},
  publisher = {American Physical Society},
  doi = {10.1103/PhysRevLett.126.141302},
  url = {https://link.aps.org/doi/10.1103/PhysRevLett.126.141302}
}

@article{Inomata,
author ={K. Inomata, Z. Lin, K. Koshino and others}, 
title={Single microwave-photon detector using an artificial $\Lambda$-type three-level system},
journal={Nat Commun},
volume={7},
pages={12303},
year={2016}
}

@article{Lescanne,
  title = {Irreversible Qubit-Photon Coupling for the Detection of Itinerant Microwave Photons},
  author = {Lescanne, Rapha\"el and Del\'eglise, Samuel and Albertinale, Emanuele and R\'eglade, Ulysse and Capelle, Thibault and Ivanov, Edouard and Jacqmin, Thibaut and Leghtas, Zaki and Flurin, Emmanuel},
  journal = {Phys. Rev. X},
  volume = {10},
  issue = {2},
  pages = {021038},
  numpages = {17},
  year = {2020},
  month = {May},
  publisher = {American Physical Society},
  doi = {10.1103/PhysRevX.10.021038},
  url = {https://link.aps.org/doi/10.1103/PhysRevX.10.021038}
}

@article{Besse,
  title = {Single-Shot Quantum Nondemolition Detection of Individual Itinerant Microwave Photons},
  author = {Besse, Jean-Claude and Gasparinetti, Simone and Collodo, Michele C. and Walter, Theo and Kurpiers, Philipp and Pechal, Marek and Eichler, Christopher and Wallraff, Andreas},
  journal = {Phys. Rev. X},
  volume = {8},
  issue = {2},
  pages = {021003},
  numpages = {9},
  year = {2018},
  month = {Apr},
  publisher = {American Physical Society},
  doi = {10.1103/PhysRevX.8.021003},
  url = {https://link.aps.org/doi/10.1103/PhysRevX.8.021003}
}

@article{Kono,
author ={S. Kono, K. Koshino, Z. Tabuchi and others}, 
title={Quantum non-demolition detection of an itinerant microwave photon},
journal={Nature Phys},
volume={14},
pages={546-549},
year={2018}
}

@article{Piedju,
author ={A. S. Piedjou Komnang and others}, 
title={Enhancing axion searches with quantum coincidence
in superconducting qubits},
doi = "10.1393/ncc/i2026-26077-4",
journal={IL NUOVO CIMENTO},
volume={49 C},
pages={77},
year={2026}
}

@Article{app14041478,
AUTHOR = {A. D’Elia and others},
TITLE = {Characterization of a Transmon Qubit in a 3D Cavity for Quantum Machine Learning and Photon Counting},
JOURNAL = {Applied Sciences},
VOLUME = {14},
YEAR = {2024},
NUMBER = {4},
ARTICLE-NUMBER = {1478},
URL = {https://www.mdpi.com/2076-3417/14/4/1478},
ISSN = {2076-3417},
DOI = {10.3390/app14041478}
}

@ARTICLE{11249713,
  author={R. Moretti and others},
  journal={IEEE Transactions on Quantum Engineering}, 
  title={Transmon Qubit Modeling and Characterization for Dark Matter Search}, 
  year={2026},
  volume={7},
  number={},
  pages={1-8},
  doi={10.1109/TQE.2025.3633176}
}

@article{DElia:2025ztz,
    author = "DElia, A. and others",
    title = "{Coherent oscillations in weakly anharmonic NbSe2 qubit}",
    eprint = "2509.17160",
    archivePrefix = "arXiv",
    primaryClass = "quant-ph",
    month = "9",
    year = "2025"
}

@article{PhysRevD.110.115021,
  title = {Search for QCD axion dark matter with transmon qubits and quantum circuit},
  author = {Chen, Shion and Fukuda, Hajime and Inada, Toshiaki and Moroi, Takeo and Nitta, Tatsumi and Sichanugrist, Thanaporn},
  journal = {Phys. Rev. D},
  volume = {110},
  issue = {11},
  pages = {115021},
  numpages = {13},
  year = {2024},
  month = {Dec},
  publisher = {American Physical Society},
  doi = {10.1103/PhysRevD.110.115021},
  url = {https://link.aps.org/doi/10.1103/PhysRevD.110.115021}
}

@article{Muck:1999ts,
    author = "M{\"u}ck, M. and Andre, M. O. and Clarke, J. and Gail, J. and Heiden, C.",
    editor = "Sikivie, P.",
    collaboration = "ADMX",
    title = "{Niobium dc SQUID with microstrip input coupling as an amplifier for the axion detector}",
    doi = "10.1016/S0920-5632(98)00517-9",
    journal = "Nucl. Phys. B Proc. Suppl.",
    volume = "72",
    pages = "145--151",
    year = "1999"
}

@article{Muck_2010,
    author = {Michael M\"uck and Robert McDermott},
    doi = {10.1088/0953-2048/23/9/093001},
    url = {https://dx.doi.org/10.1088/0953-2048/23/9/093001},
    year = {2010},
    month = {jul},
    publisher = {},
    volume = {23},
    number = {9},
    pages = {093001},
    title = {Radio-frequency amplifiers based on dc SQUIDs},
    journal = {Superconductor Science and Technology}
}

@article{bertani1999finuda,
  title={The FINUDA superconducting magnet at DA$\Phi$NE},
  author={Bertani, M and Cecchetti, A and Dulach, B and Fabbri, FL and Giardoni, M and Lanaro, A and Lucherini, V and Bressani, T and Feliciello, A and Filippi, A and others},
  doi={10.1016/S0920-5632(99)00602-7},
  journal={Nuclear Physics B-Proceedings Supplements},
  volume={78},
  number={1-3},
  pages={553--558},
  year={1999},
  publisher={Elsevier}
}

@article{ligi:2002,
  title={DA$\Phi$NE cryogenic coooling system: status and perspectives},
  author={Ligi, C and Delle Monache, G and Ricci, R and Sanelli, C},
  journal={EPAC 2002 Proceedings},
  doi={},
  volume={},
  number={},
  pages={2523--2525},
  year={2002},
  publisher={}
}

@article{Modena:1997tz,
    author = "Modena, M.",
    title = "{The DAPHNE cryogenic system}",
    reportNumber = "LNF-97-046-IR",
    year = "1997"
}

@article{Dicke,
	author = {Dicke, R. H.},
	doi = {10.1063/1.1770483},
	journal = {Rev. Sci. Instrum.},
	number = {7},
	pages = {268--275},
	title = {{The Measurement of Thermal Radiation at Microwave Frequencies}},
	volume = {17},
	year = {1946}}

@inproceedings{Gatti:2021sar,
    author = "Gatti, Claudio",
    title = "{Boosting Axion Searches with Quantum Sensing}",
    booktitle = "{14th Workshop on Low Temperature Electronics}",
    doi = "10.1109/WOLTE49037.2021.9555436",
    month = "4",
    year = "2021"
}

@article{PhysRevD.26.1817,
  title = {Quantum limits on noise in linear amplifiers},
  author = {Caves, Carlton M.},
  journal = {Phys. Rev. D},
  volume = {26},
  issue = {8},
  pages = {1817--1839},
  numpages = {0},
  year = {1982},
  month = {Oct},
  publisher = {American Physical Society},
  doi = {10.1103/PhysRevD.26.1817},
  url = {https://link.aps.org/doi/10.1103/PhysRevD.26.1817}
}

@article{Shi2023,
  title = {Ultimate precision limit of noise sensing and dark matter search},
  author = {Shi, H and Zhuang, Q},
  journal = {npj QUantum Inf},
  volume = {9},
  issue = {27},
  year = {2023},
  url = {https://doi.org/10.1038/s41534-023-00693-w}
}

@article{RevModPhys.89.035002,
  title = {Quantum sensing},
  author = {Degen, C. L. and Reinhard, F. and Cappellaro, P.},
  journal = {Rev. Mod. Phys.},
  volume = {89},
  issue = {3},
  pages = {035002},
  numpages = {39},
  year = {2017},
  month = {Jul},
  publisher = {American Physical Society},
  doi = {10.1103/RevModPhys.89.035002},
  url = {https://link.aps.org/doi/10.1103/RevModPhys.89.035002}
}

@article{Lindahl:2025zyc,
    author = "Lindahl, Jacob and Balafendiev, Rustam and Kaur, Gagandeep and Singh, Gaganpreet and Gallo Rosso, Andrea and Conrad, Jan and Gudmundsson, Jon E. and Jeong, Junu",
    title = "{Spiral tuning of wire-metamaterial cavity for a plasma haloscope}",
    eprint = "2508.18145",
    archivePrefix = "arXiv",
    primaryClass = "hep-ex",
    doi = "10.1103/n1hb-9d3p",
    journal = "Phys. Rev. Applied",
    volume = "25",
    number = "2",
    pages = "024073",
    year = "2026"
}

@article{Shandera:2018xkn,
    author = "Shandera, Sarah and Jeong, Donghui and Gebhardt, Henry S. Grasshorn",
    title = "{Gravitational Waves from Binary Mergers of Subsolar Mass Dark Black Holes}",
    eprint = "1802.08206",
    archivePrefix = "arXiv",
    primaryClass = "astro-ph.CO",
    reportNumber = "IGC-18-2-1",
    doi = "10.1103/PhysRevLett.120.241102",
    journal = "Phys. Rev. Lett.",
    volume = "120",
    number = "24",
    pages = "241102",
    year = "2018"
}

@article{LVK:2022ydq,
    author = "Abbott, R. and others",
    collaboration = "LVK",
    title = "{Search for subsolar-mass black hole binaries in the second part of Advanced LIGO{\textquoteright}s and Advanced Virgo{\textquoteright}s third observing run}",
    eprint = "2212.01477",
    archivePrefix = "arXiv",
    primaryClass = "astro-ph.HE",
    doi = "10.1093/mnras/stad588",
    journal = "Mon. Not. Roy. Astron. Soc.",
    volume = "524",
    number = "4",
    pages = "5984--5992",
    year = "2023",
    note = "[Erratum: Mon.Not.Roy.Astron.Soc. 526, 6234 (2023)]"
}

@article{Blas:2026ybh,
    author = "Blas, Diego and Chen, Yifan and Liu, Yuxin and Shang, Yanfei and Shu, Jing",
    title = "{Cavity Multimodes as an Array for High-Frequency Gravitational Waves}",
    eprint = "2601.03341",
    archivePrefix = "arXiv",
    primaryClass = "hep-ph",
    month = "1",
    year = "2026"
}

@article{Schenk:2025ria,
    author = "Schenk, Sebastian and Schmieden, Kristof and Schwaller, Pedro",
    title = "{Signatures of High-Frequency Gravitational Waves in Electromagnetic Cavities}",
    eprint = "2512.20592",
    archivePrefix = "arXiv",
    primaryClass = "hep-ph",
    reportNumber = "MITP-25-083",
    month = "12",
    year = "2025"
}

@article{Aggarwal:2025noe,
    author = "Aggarwal, Nancy and others",
    title = "{Challenges and opportunities of gravitational-wave searches above 10 kHz}",
    eprint = "2501.11723",
    archivePrefix = "arXiv",
    primaryClass = "gr-qc",
    reportNumber = "CERN-TH-2025-014, DESY-25-007",
    doi = "10.1007/s41114-025-00060-5",
    journal = "Living Rev. Rel.",
    volume = "28",
    number = "1",
    pages = "10",
    year = "2025"
}

@article{Cai:2025fpe,
    author = "Cai, Yi-fu and Visinelli, Luca and Yan, Sheng-Feng",
    title = "{Atomic Quantum Sensors for High-Frequency Gravitational Wave Searches}",
    eprint = "2510.15031",
    archivePrefix = "arXiv",
    primaryClass = "hep-ph",
    month = "10",
    year = "2025"
}

@article{Kowitt:2023wob,
    author = "Kowitt, Nolan and Balafendiev, Rustam and Sun, Dajie and Wooten, Mackenzie and Droster, Alexander and Gorlach, Maxim A. and van Bibber, Karl and Belov, Pavel A.",
    title = "{Tunable wire metamaterials for an axion haloscope}",
    eprint = "2306.15734",
    archivePrefix = "arXiv",
    primaryClass = "hep-ex",
    doi = "10.1103/PhysRevApplied.20.044051",
    journal = "Phys. Rev. Applied",
    volume = "20",
    number = "4",
    pages = "044051",
    year = "2023"
}

@article{Capdevilla:2024cby,
    author = "Capdevilla, Rodolfo and Gelmini, Graciela B. and Hyman, Jonah and Millar, Alexander J. and Vitagliano, Edoardo",
    title = "{Gravitational wave detection with plasma haloscopes}",
    eprint = "2412.14450",
    archivePrefix = "arXiv",
    primaryClass = "hep-ph",
    reportNumber = "FERMILAB-PUB-24-0436-T",
    doi = "10.1103/zn2s-2jp6",
    journal = "Phys. Rev. D",
    volume = "112",
    number = "5",
    pages = "055011",
    year = "2025"
}

@article{Schmieden:2024wqp,
    author = "Schmieden, Kristof and Schneemann, Tim and Schott, Matthias and Unni, Malavika and Bekker, Hendrik and Wickenbrock, Arne and Budker, Dmitry",
    title = "{Study of NbN as superconducting material for the usage in superconducting radio frequency cavities}",
    eprint = "2412.14958",
    archivePrefix = "arXiv",
    primaryClass = "cond-mat.supr-con",
    month = "12",
    year = "2024"
}

@article{Sathyaprakash_2009,
   title={Physics, Astrophysics and Cosmology with Gravitational Waves},
   volume={12},
   ISSN={1433-8351},
   url={http://dx.doi.org/10.12942/lrr-2009-2},
   DOI={10.12942/lrr-2009-2},
   number={1},
   journal={Living Reviews in Relativity},
   publisher={Springer Science and Business Media LLC},
   author={Sathyaprakash, B. S. and Schutz, Bernard F.},
   year={2009},
   month=mar }

@ARTICLE{1057571,
  author={Turin, G.},
  journal={IRE Transactions on Information Theory}, 
  title={An introduction to matched filters}, 
  year={1960},
  volume={6},
  number={3},
  pages={311-329},
  doi={10.1109/TIT.1960.1057571}}

@article{PhysRevD.88.035020,
  title = {Analysis of single-photon and linear amplifier detectors for microwave cavity dark matter axion searches},
  author = {Lamoreaux, S. K. and van Bibber, K. A. and Lehnert, K. W. and Carosi, G.},
  journal = {Phys. Rev. D},
  volume = {88},
  issue = {3},
  pages = {035020},
  numpages = {6},
  year = {2013},
  month = {Aug},
  publisher = {American Physical Society},
  doi = {10.1103/PhysRevD.88.035020},
  url = {https://link.aps.org/doi/10.1103/PhysRevD.88.035020}
}

@article{Brady:2022bus,
    author = "Brady, Anthony J. and Gao, Christina and Harnik, Roni and Liu, Zhen and Zhang, Zheshen and Zhuang, Quntao",
    title = "{Entangled Sensor-Networks for Dark-Matter Searches}",
    eprint = "2203.05375",
    archivePrefix = "arXiv",
    primaryClass = "quant-ph",
    reportNumber = "FERMILAB-PUB-22-131-T",
    doi = "10.1103/PRXQuantum.3.030333",
    journal = "PRX Quantum",
    volume = "3",
    number = "3",
    pages = "030333",
    year = "2022"
}

@article{Wang:2026rwy,
    author = "Wang, Yuanhong and others",
    title = "{Constraints on axion dark matter by distributed intercity quantum sensors}",
    doi = "10.1038/s41586-025-10034-w",
    journal = "Nature",
    volume = "650",
    number = "8101",
    pages = "314--319",
    year = "2026"
}

@article{Ratzinger:2024spd,
    author = "Ratzinger, Wolfram and Schenk, Sebastian and Schwaller, Pedro",
    title = "{A coordinate-independent formalism for detecting high-frequency gravitational waves}",
    eprint = "2404.08572",
    archivePrefix = "arXiv",
    primaryClass = "gr-qc",
    reportNumber = "MITP-24-040",
    doi = "10.1007/JHEP08(2024)195",
    journal = "JHEP",
    volume = "08",
    pages = "195",
    year = "2024"
}
	
\end{document}